\documentclass[12pt]{article}
\usepackage{epsfig,amsfonts}

\catcode`\@=11
\def\marginnote#1{}
\newcount\hour
\newcount\minute
\newtoks\amorpm
\hour=\time\divide\hour by60
\minute=\time{\multiply\hour by60 \global\advance\minute
by-\hour}\edef\standardtime{{\ifnum\hour<12
\global\amorpm={am}%
        \else\global\amorpm={pm}\advance\hour by-12 \fi
        \ifnum\hour=0 \hour=12 \fi
        \number\hour:\ifnum\minute<10
0\fi\number\minute\the\amorpm}}
\edef\militarytime{\number\hour:\ifnum\minute<10
0\fi\number\minute}

\def\draftlabel#1{{\@bsphack\if@filesw {\let\thepage\relax
   \xdef\@gtempa{\write\@auxout{\string
      \newlabel{#1}{{\@currentlabel}{\thepage}}}}}\@gtempa
   \if@nobreak \ifvmode\nobreak\fi\fi\fi\@esphack}
        \gdef\@eqnlabel{#1}}
\def\@eqnlabel{}
\def\@vacuum{}
\def\draftmarginnote#1{\marginpar{\raggedright\scriptsize\tt#1}}
\def\draft{\oddsidemargin -.5truein
        \def\@oddfoot{\sl preliminary draft \hfil
        \rm\thepage\hfil\sl\today\quad\militarytime}
        \let\@evenfoot\@oddfoot \overfullrule 3pt
        \let\label=\draftlabel
        \let\marginnote=\draftmarginnote

\def\@eqnnum{(\theequation)\rlap{\kern\marginparsep\tt\@eqnlabel}%
\global\let\@eqnlabel\@vacuum}  }

\def\numberbysection{\@addtoreset{equation}{section}
        \def\theequation{\thesection.\arabic{equation}}}

\def\underline#1{\relax\ifmmode\@@underline#1\else
 $\@@underline{\hbox{#1}}$\relax\fi}

\catcode`@=12
\relax
\numberbysection
\def\r2{\sqrt{2}}
\def\beq{\begin{equation}}
\def\eeq{\end{equation}}
\def\bea{\begin{eqnarray}}
\def\eea{\end{eqnarray}}
\def\beqa{\begin{eqnarray}}
\def\eeqa{\end{eqnarray}}

\def\fin{\end{document}}

\def\d(#1,#2){\delta_{#1,\, #2}}

\newcommand{\rb}{\overline{\rho}}

\newcommand{\M}{{\cal M}}
\newcommand{\ti}{\!\cdot \!}
\newcommand{\Ref}[1]{(\ref{#1})}

\begin{document}
\def\nom{}
\def\modification{}
\begin{titlepage}
\nopagebreak
\nom\modification

\begin{flushright}
LPTENS-99/54\\
hep--th/9912089
\\
December   1999
\end{flushright}

\begin{center}
{\Large
 \bf Integrable structures in classical off-shell \\ \bigskip
  10D supersymmetric Yang-Mills theory
}\\
\vglue 1 true cm
{ Jean--Loup GERVAIS\\[0.5ex]
and \\[0.5ex]
Henning SAMTLEBEN\,}\footnote{Supported by EU contract
ERBFMRX-CT96-0012.}\\ \bigskip
{\footnotesize Laboratoire de
Physique Th{\'e}orique de l'{\'E}cole Normale Sup{\'e}rieure,
\footnote{UMR 8548:  Unit{\'e} Mixte du Centre National de la Recherche
Scientifique, et de
l'{\'E}cole Normale Sup{\'e}rieure. }\\ 24 rue Lhomond, 75231 Paris
C{\'E}DEX 05, ~France.}
\end{center}

\baselineskip .4 true cm
\noindent
\begin{abstract}
The superspace flatness conditions which are equivalent to 
the field equations of supersymmetric Yang-Mills theory in ten
dimensions have not been useful so far to derive non trivial
classical solutions.  Recently, modified flatness conditions were
proposed, which are explicitly integrable  \cite{GerSav99}, and are
based on the breaking of symmetry $SO(9,1)\to SO(2,1)\otimes SO(7)$.  In
this article, we investigate their physical content. To this end,
group-algebraic methods are developed which allow to derive the set of
physical fields and their equations of motion from the superfield
expansion of the supercurl, systematically.

A set of integrable superspace constraints is identified which
drastically reduces the field content of the unconstrained superfield
but leaves the spectrum including the original Yang-Mills vector field
completely off-shell. A weaker set of constraints gives rise to
additional fields obeying first order differential equations.
Geometrically, the $SO(7)$ covariant superspace constraints descend
from a truncation of Witten's original linear system to particular
one-parameter families of light-like rays.
\end{abstract}
\vfill

\end{titlepage}
\tableofcontents

\newpage

 \section{Introduction.}

Recently, progress was made in applying exact integration methods to
supersymmetric Yang-Mills theory in ten dimensions starting from
 the flatness conditions in superspace which have been known for
some time to be equivalent to the field equations
\cite{Nils81,Witt86}. It was shown in \cite{GerSav99}, that there
exists an on-shell light cone gauge, where the superfields may be
entirely expressed in terms of a scalar superfield satisfying two sets
of equations. The first is linear and the general solution was derived;
the second is similar to Yang's equations and has been handled by
methods similar to earlier studies of self-dual Yang-Mills in four
dimensions. A general class of exact solutions\footnote{keeping,
however, only the dependence upon one time and one space coordinates
in contrast with the dimensional reduction which will be discussed
below. This is probably not essential. } has been obtained
\cite{GerSav99} and a B{\"a}cklund transformation put forward
\cite{Gerv99c}.

So far, however, it has not been possible to simultaneously solve the
two sets of equations.  Only a particular class of solutions of the
Yang type equations has been found, which is not general enough to
solve the other (linear) set. Returning to a general gauge, one may
see that deriving the scalar superfield satisfying the linear subset
of equations is equivalent \cite{Gerv99c} to solving a particular
symmetrized form of the flatness conditions. This symmetrized form was
shown to be explicitly integrable directly, since it arises
\cite{Gerv99a,Gerv99b} as compatibility condition of a Lax
representation, similar to the one of Belavin and Zakharov
\cite{BelZak78a}, which may be handled by the same powerful techniques
as in the case of self-dual Yang-Mills in four dimensions.
\medskip

The goal of this article is a systematic study of these integrable
modifications of the original flatness conditions in superspace. We shall
refer to them as the {\em integrable} superspace constraint -- as opposed
to the {\em strong} superspace constraint which describes vanishing of
the full super field strength and is equivalent to the Yang-Mills system.
Since the proof of this equivalence is recursive and rather tedious
\cite{Nils81,Witt86,HarShn86,AbFoJa88}, a priori it is not clear which
modification of the field content and dynamics is implied by a
modification of the strong superspace constraint.

An important property of the integrable superspace constraint is that
it explicitly breaks the original $SO(9,1)$ Lorentz symmetry down to
$SO(2,1)\times SO(7)$. For Weyl-Majorana spinors and in particular the
Grassmann coordinates in superspace, this leads to the separation
${\bf 16} \rightarrow {\bf 8}+{\bf 8}$, (where the r.h.s.\ denotes a
pair of $SO(7)$ spinor representations which form a doublet under
$SO(2,1)$). This is instrumental in defining the involution which is
the key to applying methods modeled over the bosonic self-duality
requirement in four dimensions; here, the analogous construction
involves the exchange of the two spinor representations.

In the rest of the paper we concentrate upon deriving the field
content and dynamical equations induced by the integrable superspace
constraint. Since this constraint is weaker than the original (strong)
superspace constraint, we effectively go partially off-shell. 
There are more physical fields in the superfield components
and  the dynamics is modified. At this point, it is worth recalling that,
in the standard treatment~\cite{HarShn86,AbFoJa88}, the method used to
eliminate the unphysical components of the superfields makes an essential
use of the equations of motion. It is thus not applicable to our case.
One of the aims of the present work is to devise a more direct and general
method, which is also applicable to our modified equations.

\medskip

After reviewing superspace notations, in section~2 we introduce the
original (strong) and the modified (integrable) superspace
constraints.  We explain the geometrical origin of the integrable
superspace constraint as particular truncation of the Lax
representation in superspace \cite{Witt86}. This original Lax
representation, associated with the strong superspace constraint, is
formulated for light-like rays in ten dimensions, which play the role
of spectral parameters. Restricting the connection to certain
one-parameter families of light rays (spanning a three-manifold) makes
the system accessible to the techniques developed in \cite{BelZak78a} for
four-dimensional self-dual Yang-Mills theory. With a particular choice
of three-manifold, this gives back the integrable superspace
constraint. For the subsequent analysis it is helpful to further
introduce a slightly stronger version of this constraint corresponding
to an effective reduction of the Lax representation to seven
dimensions and referred to as the {\em reduced integrable}
constraint. Due to the fact that this reduced constraint commutes with
the action of the symmetric group on spinor indices, it eventually
turns out to be completely soluble, which essentially simplifies the
analysis of the integrable constraint.

The rest of the paper is devoted to studying field content and
dynamics induced by the integrable and the reduced integrable
superspace constraints.  Section~3 presents a systematic general study
of the expansion of the superfield equations in powers of the odd
variables $\theta$. We derive recursion relations with an interesting
structure. The elimination of unphysical components of superfields is
done recursively and involves two operators noted $S$ and $T$. The
first satisfies a simple quadratic equation while the second is
nilpotent. Thus our equations bear some analogy with the descent
equations \cite{Zumi85}. The field equations are enforced by further
applying a projector $K$, and we thus study the interplay between $S$,
$T$, and $K$ on general ground.

Applying this method to the integrable superspace constraints, in
section~4, we explicitly identify the induced physical field
content. As a result, we find a spectrum which is essentially larger
than the original Yang-Mills system. For the reduced integrable
constraint it consists of $384\!+\!384$ fields which correspond to
three copies of the ten-dimensional off-shell multiplet
\cite{BerRoo82}; the integrable constraint then gives rise to
additional $31\!+\!16$ fields. Section~5 is devoted to deriving
explicit recurrent relations which determine the higher order
superfield components in terms of these fields.

Finally, in section~6 we derive the field equations which are induced
by the superspace constraints. The spectrum associated with the
reduced integrable constraint remains completely off-shell, whereas
the additional fields appearing with the integrable constraint satisfy
a set of first order differential equations. A discussion of this
dynamics and some concluding remarks are given at the end.

\section{Superspace constraints and Lax representation.}

\subsection{Superfield conventions}

The notations are essentially the same as in the previous references.
The physical fields are noted as follows: $X_a(\underline x)$ is the
vector potential, $\phi^\alpha(\underline x) $ is the Majorana-Weyl
spinor. Both are matrices in the adjoint representation of the gauge
group ${\bf G}$. Latin indices $a=0,\ldots, 9$ describe Minkowski
components, Greek indices $\alpha=1,\ldots, 16$ denote spinor
components. We use the Dirac matrices
\beq
\Gamma^a=\left(\begin{array}{cc}
0_{16\times 16}&\left(\left(\sigma^a\right)^{\alpha\beta}\right)\\
\left(\left(\sigma^a\right)_{\alpha\beta}\right)&0_{16\times 16}
\end{array}\right),\quad
\Gamma^{11}= \left(\begin{array}{cc}
1_{16\times 16}&0\\0&-1_{16\times 16}\end{array}\right)\;.
\label{real1}
\eeq
We will use the superspace formulation with 16 odd coordinates
$\theta^\alpha$. The general superfield expansions of a superfield
$\Phi\left(\underline x, \underline \theta \right)$ may be written as
\beq
\Phi(\underline x, \underline \theta)~=~
\sum_{p=0}^{16}
\Phi^{[p]}(\underline x,\, \underline \theta ) ~\equiv~
\sum_{p=0}^{16}
\sum_{\alpha_1,\ldots,
\alpha_p} {\theta^{ \alpha_1}\cdots
\theta^{ \alpha_p}\over p !}\,
\Phi^{[p]}_{\alpha_1\ldots \alpha_p}(\underline x)\;.
\label{exp}
\eeq
The grading is given by the operator
\beq
{\cal R}=\theta^\alpha\partial_\alpha \;,\qquad
\left[ \,{\cal R}\,,\Phi^{[p]} \,\right] = p\,\Phi^{[p]} \;.
\label{Rdef}
\eeq
Superderivatives are defined by
\beq
D_\alpha=\partial_\alpha-\theta^\beta
\left(\sigma^a\right)_{\alpha \beta}  {\partial_a}\;, \quad
\mbox{such that}\quad
\left[D_\alpha\,,D_\beta \,\right]_+ =
-2\,(\sigma^a)_{\alpha \beta}
\,\partial_a  \;.
\label{susy}
\eeq
The odd super vector potential, valued in the adjoint representation
of the gauge group, is denoted by $A_\alpha\left(\underline
x,\underline \theta\right)$. We define its supercurl $M_{\alpha\beta}$
as
\bea
M_{\alpha\beta}&\equiv& D_\alpha A_\beta+D_\beta A_\alpha+
\left[A_\alpha, A_\beta\right]_+ ~\equiv~
F_{\alpha \beta}-
2\left(\sigma^a\right)_{\alpha\beta}A_a\;,  \label{curdef} \\[1ex]
A_a &\equiv&
-{1\over32}\, (\sigma_a)^{\alpha\beta}\,M_{\alpha\beta}\;.
\label{Adef}
\eea
This gives the decomposition of the supercurl into the super field
strength $F_{\alpha \beta}$ and the even vector potential
$A_a\left(\underline x,\underline \theta\right)$ as functions of
$A_\alpha$.\footnote{In the terminology of \cite{GaStWe80}, we have hence
resolved the ``conventional constraint''
$(\sigma_a)^{\alpha\beta}F_{\alpha \beta}=0$.}  The superfield
formalism is invariant under gauge transformations
\bea
A_\alpha &\mapsto& g^{-1}A_\alpha\,g + g^{-1}D_\alpha g \;,
\label{sugauge} \\
F_{\alpha\beta} &\mapsto& g^{-1}F_{\alpha\beta}\,g \;,\nonumber\\
A_a &\mapsto& g^{-1}A_a\,g + g^{-1}\partial_a g  \;,\nonumber
\eea
with an even superfield $g\left(\underline x,\underline \theta\right)$
as gauge parameter. Imposing the so-called recursion gauge condition
\beq
\theta^\alpha A_\alpha=0
\label{supgge}
\eeq
restricts the  freedom \Ref{sugauge} to ordinary gauge
freedom, i.e.~to gauge parameters $g$ with $[{\cal{R}},g]=0$.

\subsection{Superspace constraints}

It is known that the equations of motion of super Yang Mills theory in
ten dimensions may equivalently be expressed as vanishing of the super
curvature
\beq\label{flat}
F_{\alpha \beta}~=~0\;.
\eeq
More precisely, it has been shown in
\cite{Nils81,Witt86,HarShn86,AbFoJa88} that the recursion gauge
condition \Ref{supgge}
together with the flatness conditions \Ref{flat} implies the
Yang-Mills equations of motion
\bea
\partial^a F_{ab} + \left[\,A^a\,,F_{ab}\,\right] &=&
{1\over2}\, (\sigma_b)_{\alpha\beta}\,
\left[\,\chi^\alpha\,,\chi^\beta\,\right] \label{YM}\\[1ex]
(\sigma^a)_{\alpha\beta}\,
\left(\partial_a\chi^\beta + \left[\,A_a\,,\chi^\beta\,\right]\right)
&=& 0 \;, \nonumber
\eea
for the superfields $A_a$ and $\chi^\alpha$, the latter being defined
as $\chi^\alpha \equiv (\sigma^a)^{\alpha\beta}\,F_{a\beta}$ with the
curvature
\beq\label{Faa}
F_{a\beta} \equiv
\left(D_\beta A_a - \partial_a A_\beta +
\left[\,A_\beta\,,A_a\,\right]\;
\right) \;.
\eeq
Moreover, \Ref{supgge} and \Ref{flat} yield a unique recurrent
prescription of the higher order superfield components in $A_a$,
$\chi^\alpha$ and $A_\alpha$ as functions of the zero order
contributions
\beq\label{physical}
X_a \equiv A_a^{[0]}\;,\qquad \phi^\alpha \equiv \chi^{\alpha\,[0]}\;,
\eeq
These zero order components in particular satisfy the usual
supersymmetric Yang-Mills equations of motion.

For the purpose of this paper we rewrite the vanishing super curvature
condition \Ref{flat} in terms of the supercurl $M_{\alpha\beta}$. It is
convenient to introduce the general space of superfields symmetric in
two additional spinor indices $(\alpha\beta)$, which we denote by
$\M$. This space carries the grading \Ref{exp}:
\beq
\M ~= ~ \bigoplus_{p=0}^{16} \M^{[p]} ~\equiv~
\bigoplus_{p=0}^{16}
\left\{ v_{(\alpha\beta),[\gamma_1\dots\gamma_p]}\;
        \theta^{\gamma_1}\dots\theta^{\gamma_p} \right\} \;.
\label{cM}
\eeq
and its elements have the general decomposition according to
\bea
\underbrace{M_{\alpha\beta}}_{({\bf 16}\!\times \!{\bf 16})_s} &=&
\underbrace{-2 \left(\sigma^a\right)_{\alpha \beta}A_a}_{\bf 10}~+~
\underbrace{{\textstyle{1\over 5!}}
\left(\sigma^{a_1\ldots a_5}\right)_{\alpha
\beta}B_{{a_1\ldots a_5}}}_{\bf 126}\;, \label{Mdec}
\eea
with selfdual $B_{{a_1\ldots a_5}} = -{\textstyle{1\over 5!}}\,
\epsilon_{a_1\ldots a_{10}}\,B^{a_6\ldots a_{10}}$.
In terms of this decomposition, the superspace constraint \Ref{flat}
corresponds to setting $B_{{a_1\ldots a_5}}= 0$ and may be written as
a projection condition $K_{{\rm \scriptstyle YM}}$ on the supercurl
\bea
&&M_{\alpha\beta}~=~(K_{{\rm \scriptstyle YM}}\,M)_{\alpha\beta}\;,
\label{con10}\\[2ex]
&&\mbox{with }\quad
(K_{{\rm \scriptstyle YM}})_{\alpha \beta}^{\alpha'
\beta'}={\textstyle{1\over 16}}
\left(\sigma^a\right)_{\alpha\beta}
\left(\sigma_a\right)^{\alpha' \beta'}\;,\nonumber
\eea

This paper is devoted to a study of other (weaker) superfield
constraints which replace \Ref{con10} and have appeared as completely
integrable superfield equations in \cite{Gerv99a,Gerv99b}. Obviously,
\Ref{con10} is the only $SO(9,1)$ covariant constraint that can be
imposed on the supercurl. We hence break the original $SO(9,1)$
Lorentz invariance of the system down to $SO(2,1)\times SO(7)$; the
reason for this particular choice will become clear the following. It
is in this setting that the modified constraints have appeared in
\cite{GerSav99,Gerv99a,Gerv99b}.

For Weyl-Majorana spinors, this symmetry breaking leads to the
separation
\beq
{\bf 16} \rightarrow ({\bf 2},{\bf 8})\;:\quad
\chi^{\alpha}\mapsto (\chi^{\mu},\chi^{\bar\mu})\;,\quad
\mu,\bar\mu=1, \dots, 8\;,
\eeq
where the r.h.s.\ denotes a pair of $SO(7)$ spinor representations
which form a doublet under $SO(2,1)$ such that
$\mu\leftrightarrow\bar\mu$ denotes the $SO(7)$ invariant
involution. In appendix~A we have collected the conventions about
decomposing the $SO(9,1)$ $\sigma$-matrices into $SO(8)$
$\gamma$-matrices.

The supercurl $M_{\alpha\beta}$ correspondingly decomposes into
\beq
M_{\alpha\beta}~=~{\bf 10}\!+\!{\bf 126}~\mapsto~
({\bf 3},{\bf 1}) + ({\bf 1},{\bf 7})
+({\bf 1},{\bf 21}) + ({\bf 3},{\bf 35})
\eeq

The superspace constraints which we are going to study in this paper
are the following projections
\bea
K_{{\rm \scriptstyle YM}}:&&
M_{\alpha\beta}~\rightarrow~
({\bf 3},{\bf 1}) + ({\bf 1},{\bf 7})\label{KYM}\\[1ex]
K_{{\rm \scriptstyle I}}:&&
M_{\alpha\beta} ~\rightarrow~
({\bf 3},{\bf 1}) + ({\bf 1},{\bf 7})+({\bf 1},{\bf 21})
\label{KI}\\[1ex]
K_{{\rm \scriptstyle IR}}:&&
M_{\alpha\beta}~\rightarrow~
({\bf 1},{\bf 7})+({\bf 1},{\bf 21})
\label{KIR}
\eea
The first constraint is the original vanishing super curvature
condition \Ref{con10} which corresponds to the Yang-Mills system. The
latter two constraints have appeared in \cite{Gerv99a,Gerv99b} as
compatibility equations of completely integrable Lax
representations. We will refer to them as to the integrable and the
reduced integrable constraint, respectively. Note, that the truncation
\Ref{KIR} is gauge covariant only after dimensional reduction of the
system to seven dimensions.

Before analyzing the field content and dynamics implied by these
superfield constraints we first recall how they may be obtained as
compatibility equations of Lax representations in superspace.

\subsection{Lax representations}

In this section we recall the original linear system \cite{Witt86}
associated with the vanishing curvature condition \Ref{KYM}, and show
how the integrable constraints \Ref{KI}, \Ref{KIR} may be obtained as
certain truncations thereof.

The flatness conditions \Ref{flat}, \Ref{KYM} possess a Lax type
representation in superspace \cite{Witt86}. They imply the existence of a
${\bf G}$-valued superfield $\Psi[\underline\ell]$ for any light-like
ten-dimensional vector $\underline \ell$, which is defined by the
linear system
\bea
\ell^a \left(\sigma_a\right)^{\alpha \beta}
\left\{ D_\beta + A_\beta \right\} \,\Psi[\underline \ell] &=& 0 \;,
\label{flat1}\\[1ex]
\ell^a \left\{\partial_a -  A_a \right\}
\,\Psi[\underline \ell] &=& 0 \;.
\nonumber
\eea
In turn, the compatibility conditions of \Ref{flat1} imply
\Ref{flat}. Clearly, these equations are invariant under
multiplication by an overall constant, so that $\Psi$ only depends
upon the light-like ray considered.

Equations \Ref{flat1} may be considered as a Lax representation of the
field equations \Ref{YM} where the light-like vector $\underline \ell$
plays the role of the spectral parameter.  As they stand, however,
they have not been very useful in practice, i.e.\ with regard to the
powerful solution generating methods applicable in lower dimensional
systems. The constraints \Ref{KI} and \Ref{KIR}, in contrast, are
derivable from a Lax representation where the spectral parameter is a
complex number $\lambda$ instead of a light ray, such that methods
inspired from Ref.~\cite{BelZak78a} become applicable
\cite{Gerv99b,Gerv99c}. Starting from equations \Ref{flat1}, the
integrable constraints geometrically correspond to keeping only the
flatness conditions associated with a particular one parameter subset
$\underline \ell (\lambda) $ of light-like rays.

Breaking the original $SO(9,1)$ invariance we introduce the following
parametrization for light-like rays $\ell=\ell(\lambda,\vec v)$:
\beq\label{lam}
\ell^\pm = \pm i\,{1\mp\lambda\over 1\pm\lambda } \;,\quad
\ell^i = v^i\;,\quad\mbox{for}\quad i=1,\dots,8 \;,
\quad\ell^\pm \equiv \ell^0\pm \ell^9 \;,
\eeq
with a complex number $\lambda$ and an eight-dimensional unit vector
$v^i$. The vector $v^i$ provides a mapping between the two spinor
representations of $SO(8)$ via $\gamma^{\vec v}_{\mu\bar\nu}=\vec
v\cdot\vec\gamma_{\mu\bar\nu}$ (cf.~appendix~A).

The linear system \Ref{flat1} then takes the form
\bea\label{BZ}
\left\{\Delta^{\vec v}_\mu +B^{\vec v}_\mu
+\lambda (\overline{\Delta}^{\vec v}_{ \mu }+\overline{B}^{\vec v}_\mu )
\right\} \Psi [\lambda,\vec v]&=& 0\;,\\[2ex]
\left\{\frac{1\!-\!\lambda}{1\!+\!\lambda}\,(\partial_+\!+\!A_+) -
\frac{1\!+\!\lambda}{1\!-\!\lambda}\, (\partial_-\!+\!A_-)
 -i\,\vec v\!\cdot\!(\vec\partial\!+\!\vec A)
\right\} \Psi [\lambda,\vec v] &=&0\;,
\nonumber
\eea

\noindent
with
\bea
\Delta^{\vec v}_\mu &=&D_\mu +i \gamma^{\vec v}_{\mu \overline
\rho}D_{\overline \rho}\;,\quad
\overline{\Delta}^{\,\vec v}_\mu ~=~D_\mu -i \gamma^{\vec v}_{\mu
\overline \rho}D_{\overline \rho} \;,\label{Lv3}\\
B^{\vec v}_\mu &=&A_\mu
+i \gamma^{\vec v}_{\mu \overline \rho}A_{\overline \rho}\;,\quad
\overline{B}^{\,\vec v}_\mu ~=~
A_\mu -i \gamma^{\vec v}_{\mu \overline \rho}A_{\overline \rho} \;.
\nonumber
\eea

We are now going to show, that the superspace constraints \Ref{KYM},
\Ref{KI}, and \Ref{KIR} arise as compatibility equations of certain
truncations of \Ref{BZ}:
\bea
K_{{\rm \scriptstyle YM}}:&&
\mbox{impose \Ref{BZ} for all vectors $\vec v$} \label{KKK}\\[1ex]
K_{{\rm \scriptstyle I}}:&&
\mbox{impose \Ref{BZ} for a fixed vector $\vec v$} \nonumber\\[1ex]
K_{{\rm \scriptstyle IR}}:&&
\mbox{impose \Ref{BZ} for a fixed vector $\vec v$}\nonumber\\
&&\mbox{and reduce the linear system to seven dimensions}
\nonumber
\eea

The first relation is the result of \cite{Witt86} and follows from
computing the commutator of the Lax connection \Ref{BZ} with itself,
thereby implying
$$
(1\!+\!\lambda)^2\,F_{\mu \nu} -
(1\!-\!\lambda)^2\,
\gamma^{\vec v}_{\mu\bar\mu}\gamma^{\vec v}_{\nu\bar\nu}\,
F_{\overline\mu  \overline{\nu}} + i(1\!-\!\lambda^2)\,
(\gamma^{\vec v}_{\mu\bar\nu}F_{\nu \overline \nu}
+\gamma^{\vec v}_{\nu\bar\mu}F_{ \mu \overline \mu }) ~=~0
$$
for all values of $\lambda$ and $\vec v$ and hence vanishing of the
supercurvature \Ref{flat}.

For a fixed choice of the vector $\vec v$ on the other hand, these
conditions imply
\begin{eqnarray}
M_{\mu\nu} &=& 4\,\delta_{\mu \nu}A_+~=~ {\textstyle\frac18}\,
\delta_{\mu\nu}\,\delta^{\mu'\nu'}\,M_{\mu'\nu'}\;,
\label{dynuu}\\[1ex]
M_{{\overline \mu} {\overline \nu}} &=&
4\,\delta_{{\overline\mu }{\overline\nu}}A_-~=~
{\textstyle\frac18}\,
\delta_{\overline\mu\overline\nu}\,
\delta^{\overline\mu'\overline\nu'}\,M_{\overline\mu'\overline\nu'}\;,
\nonumber\\[1ex]
(\gamma^{\vec v}_{\mu\bar\nu}M_{\nu \overline \nu}
+\gamma^{\vec v}_{\nu\bar\mu}M_{ \mu \overline \mu }) &=&
-2\,\delta_{\mu\nu}\,\vec v\!\cdot\!\vec A ~=~
{\textstyle\frac14}\,\delta_{\mu\nu}\,
\gamma^{\vec v\,\,}{}^{\mu'\bar\nu'}\,M_{\mu'\bar\nu'}\;.
\nonumber
\end{eqnarray}
This precisely corresponds to the projection \Ref{KI}.
If furthermore we assume independence of the solution $\Psi$ of the
Lax pair \Ref{BZ} of the three coordinates
\beq\label{drop}
\partial_\pm\Psi~=~0~=~\vec v\!\cdot\!\vec\partial\,\Psi\;,
\eeq
the second equation of \Ref{BZ} shows that
\bea
A_\pm &=&  \Psi\;\partial_\pm \Psi^{-1} \Big|_{\lambda=\mp1} ~=~0\;,
\label{AP1}\\[1ex]
\vec v\!\cdot\!\vec A &=&  \nonumber
\Psi\;(\vec v\!\cdot\!\vec\partial+i\partial_9) \Psi^{-1}
\Big|_{\lambda=0}-iA_9 ~=~0\;.
\nonumber
\eea
Together with \Ref{dynuu}, this implies the following stronger
truncation on the supercurl
\bea
M_{\mu\nu} &=& 0\;,
\label{dynuur}\\[1ex]
M_{{\overline \mu} {\overline \nu}} &=&0\;,
\nonumber\\[1ex]
(\gamma^{\vec v}_{\mu\bar\nu}M_{\nu \overline \nu}
+\gamma^{\vec v}_{\nu\bar\mu}M_{ \mu \overline \mu }) &=&0\;,
\nonumber
\eea
corresponding to the projection \Ref{KIR}. This finishes the proof of
\Ref{KKK}. We have hence shown, that the superspace constraints
\Ref{KI} and \Ref{KIR} arise as integrability conditions of the Lax
pair \Ref{BZ} upon truncating the spectral parameter to a particular
one parameter family of light-like rays.

The advantage of the reduced Lax connection comes from the fact that
for fixed choice of $\vec v$, the linear system \Ref{BZ} is similar to
the one proposed by Belavin and Zakharov for four-dimensional
self-dual Yang-Mills theory \cite{BelZak78a} and similar techniques may
successfully be applied. The role of the involution which in that case
describes self-duality is played by exchanging the two $SO(8)$ spinor
representations by means of $\vec v\cdot \vec \gamma_{\mu \overline
\rho}$, here. This explains the particular choice of breaking the
original Lorentz symmetry down to $SO(9,1)\rightarrow SO(2,1)\times
SO(7)$. Upon this reduction, the system \Ref{BZ} may be solved
starting from an ansatz which is meromorphic in $\lambda$. This leads
to purely algebraic equations -- coming from the fact that the bracket
in equation \Ref{BZ} is linear in $\lambda$ -- which may be solved in
essentially the same way as was done for the self-dual Yang-Mills
theory. In this sense, the constraints \Ref{KI} and \Ref{KIR} arise as
completely integrable superfield equations.

In contrast, it seems impossible to carry out the next step and solve
the system for all $\vec v$, which would really give a solution of the
full Yang-Mills equations in ten dimensions. Indeed, equation
\Ref{Lv3} implies that the bracket in \Ref{BZ} should be linear in
$\vec v.\vec \gamma$, a very strong requirement, which to satisfy
there seems to exists no systematic method.

The situation further simplifies upon dropping the coordinate
dependence according to \Ref{drop}. It should be noted that this is a
natural but stronger requirement than solely restricting the
coordinate dependence of the superfield components of
$M_{\alpha\beta}$ (cf.\ \cite{ChaMil87} for a discussion of the
reduction to four dimensions). With the original linear system
\Ref{flat1} for example, dimensionally reduced physical configurations
generically induce functions $\Psi$ which still depend on the
compactified coordinates. Restricting to dimensionally reduced
functions $\Psi$ corresponds to imposing further superfield
constraints \Ref{dynuur}.

Relaxing the original superspace constraints however corresponds to
going partially off-shell and gives rise to additional fields
appearing in the higher order superfield components. Our goal in the
following is to extract the field and the dynamical content associated
with the integrable subset of superspace constraints.

\bigskip

\section{Systematics of the supercurl expansion.}

In the following, we study the purely algebraic problem to determine
the field and dynamical content, induced by imposing the constraints
\Ref{KYM}--\Ref{KIR} on the superfields. To this end, we first review
the level structure of the superfields and show how to systematically
extract field content and equations.

For the original set of constraints \Ref{dynuu} this has been
discussed in detail in \cite{HarShn86,AbFoJa88} (and likewise in
\cite{HHLS85} for the reduction to $N\!=\!3$ supersymmetric Yang-Mills
theory in four dimensions).  However, this discussion makes an
essential use of the Yang-Mills field equations, and thus does not
apply to our case.  The purpose of this paper is to present an
alternative method. As central object we consider the supercurl
$M_{\alpha\beta}$.

\subsection{Algebraic structure of the recursion gauge condition}

If we do not impose any constraint on the superfields, we simply have
to take into account the fact that the gauge freedom \Ref{sugauge} has
been fixed by the recursion gauge condition \Ref{supgge} to gauge
parameters which do not have higher order superfield components. For
the components of the super vector potential this implies
\beq\label{supgec}
\theta^\alpha A_\alpha=0 \qquad \Longleftrightarrow \qquad
A^{[p]}_{[\alpha,\gamma_1\dots \gamma_p]} =0\;,
\eeq
which is still invariant under ordinary gauge transformations.

This shows that the independent components in the superfield
$A_\alpha$ are given by the following sum of Young
diagrams\footnote{Here, and in the following, the notion of Young
diagrams always refers to (anti)symmetrizing the factors of a tensor
product $V^{\otimes N}$ of a given representation $V$, i.e.\ always to
the Young diagrams of the corresponding permutation group ${\mathfrak
S}_N$.} for the spinor representation ${\bf 16}$ of $SO(9,1)$

\beq\label{young}
\eeq
\vspace{-3.5em}
\begin{figure}[htbp]
  \begin{center}
    \leavevmode
\begin{picture}(0,0)%
\epsfig{file=young.pstex}%
\end{picture}%
\setlength{\unitlength}{0.00083300in}%
\begingroup\makeatletter\ifx\SetFigFont\undefined%
\gdef\SetFigFont#1#2#3#4#5{%
\reset@font\fontsize{#1}{#2pt}%
\fontfamily{#3}\fontseries{#4}\fontshape{#5}%
\selectfont}%
\fi\endgroup%
\begin{picture}(4024,1369)(889,-1498)
\put(4426,-261){\makebox(0,0)[lb]{\smash{\SetFigFont{11}{13.2}
{\rmdefault}{\mddefault}{\updefault}$A_\alpha^{[p]}$}}}
\put(1389,-502){\makebox(0,0)[lb]{\smash{\SetFigFont{11}{13.2}
{\rmdefault}{\mddefault}{\updefault}$+$}}}
\put(2213,-502){\makebox(0,0)[lb]{\smash{\SetFigFont{11}{13.2}
{\rmdefault}{\mddefault}{\updefault}$+$}}}
\put(3038,-502){\makebox(0,0)[lb]{\smash{\SetFigFont{11}{13.2}
{\rmdefault}{\mddefault}{\updefault}$+\quad\dots\quad +$}}}
\put(4913,-502){\makebox(0,0)[lb]{\smash{\SetFigFont{11}{13.2}
{\rmdefault}{\mddefault}{\updefault}$+\quad\dots$}}}
\put(4404,-1178){\makebox(0,0)[lb]{\smash{\SetFigFont{11}{13.2}
{\rmdefault}{\mddefault}{\updefault}$\vdots$}}}
\put(4020,-960){\makebox(0,0)[lb]{\smash{\SetFigFont{10}{12.0}
{\rmdefault}{\mddefault}{\updefault}$p
\left\{\vphantom{\begin{array}{c} \\[13ex]\end{array}} \right.$}}}
\put(2626,-261){\makebox(0,0)[lb]{\smash{\SetFigFont{11}{13.2}
{\rmdefault}{\mddefault}{\updefault}$A_\alpha^{[3]}$}}}
\put(976,-261){\makebox(0,0)[lb]{\smash{\SetFigFont{11}{13.2}
{\rmdefault}{\mddefault}{\updefault}$A_\alpha^{[1]}$}}}
\put(1801,-261){\makebox(0,0)[lb]{\smash{\SetFigFont{11}{13.2}
{\rmdefault}{\mddefault}{\updefault}$A_\alpha^{[2]}$}}}
\end{picture}
\end{center}
\end{figure}

\paragraph{Recurrence relations }
Since later on we are going to study the field equations implied by
further constraining the supercurl $M_{\alpha \beta}$, we first
identify the remaining independent field components in $M_{\alpha
\beta}$ after imposing the recursion gauge condition. Equation
\Ref{supgec} yields
\beq
\left(1+{\cal R}\right) A_\alpha=\theta^\beta M_{\alpha\beta}
\qquad \Longleftrightarrow \qquad
A_\alpha^{[p+1]}={1\over p+2}\,\theta^\gamma M^{[p]}_{\alpha \gamma}\;.
\label{AM}
\eeq
\newpage

\noindent
At order $p$ we get from \Ref{curdef}
\bea
M^{[p]}_{\alpha \beta}&=&
\partial_\alpha A^{[p+1]}_{\beta }+ \partial_\beta
A^{[p+1]}_{\alpha } \nonumber\\
&&{}-\left(\sigma^m\right)_{\alpha \gamma}\theta^\gamma
\partial_m A^{[p-1]}_\beta
-\left(\sigma^m\right)_{\beta \gamma}\theta^\gamma
\partial_m A^{[p-1]}_\alpha+
\sum_{q=1}^{p-1}\left[A^{[q]}_\alpha,\, A^{[p-q]}_\beta\right]_+
\nonumber
\eea
Using \Ref{AM} we may then re-express this relation entirely in terms
of $M_{\alpha\beta}$. It is convenient to write it in the form
\medskip
\beq
\left(S+{\cal R}\right)\,M+{{\cal R}+2\over {\cal R}}\;T\,M~=~{\cal C}
\;,
\label{geneq}
\eeq
\medskip

\noindent
where we have introduced two linear operators on the set of symmetric
superfields, by $\left(SM\right)_{\alpha \beta}=S_{\alpha
\beta}^{\alpha'\beta'}M_{\alpha' \beta'} $, and
$\left(TM\right)_{\alpha \beta}=T_{\alpha
\beta}^{\alpha'\beta'}M_{\alpha' \beta'} $, with
\bea
S_{\alpha \beta}^{\alpha'\beta'}&=&\delta_{\alpha}^{\alpha'}
\theta^{\beta'}\partial_\beta+
\delta_{\beta}^{\beta'}\theta^{\alpha'}\partial_\alpha
\label{Sdef}\\[4pt]
T_{\alpha \beta}^{\alpha' \beta'}&=&
\theta^\gamma \left(
 \theta^{\beta'} \left(\sigma^a\right)_{\beta\gamma}
 \delta_{\alpha}^{\alpha'}
+ \theta^{\alpha'} \left(\sigma^a\right)_{\alpha \gamma}
 \delta_{\beta}^{\beta'}
  \right)\partial_a
\label{Tdef}
\eea
Moreover, the non linear term ${\cal C}$ is given by
\beq
{\cal C}^{[p]}_{\alpha \beta}=\left(p+2\right)
\sum_{q=1}^{p-1}{\left[
\theta^\gamma M^{[q-1]}_{\alpha \gamma}
,\,
\theta^\delta M^{[p-q-1]}_{\beta \delta}
\right]_+ \over \left(q+1\right)\left(p-q+1\right)} \;.
\label{Cdef}
\eeq

Note, that the operator $S$ commutes with ${\cal R}$ whereas $T$
raises the level by 2.
Thus, \Ref{geneq} indeed builds a recursive
system, relating the higher levels of $M_{\alpha\beta}$ to the image
of the lower ones under $T$.
\bigskip

\mathversion{bold}
\paragraph{Algebraic properties of $S$ and $T$}
\mathversion{normal}

By explicit computation one verifies that the operator $S$ satisfies
the equation
\beq
\left(S-2\right)\left(S+{\cal R}\right)=0
\label{Seq}
\eeq

\noindent
Thus, at a given level ${\cal R}=p$, the operator $S$ has only two
different eigenvalues. We may hence decompose $\M$ into the
eigenspaces of $S$:
\beq\label{Fpmdef}
\M=\M^++\M^-\;,\qquad
\left(S+{\cal R}\right)\,\M^+=0\;,\quad
\left(S-2\right)\M^- =0 \;.
\eeq
With \Ref{Sdef} one finds that $(S\!-\!2)$ and $(S\!+\!{\cal R})$
are proportional to the projectors onto the following Young diagrams

\beq\label{M+M-}
\eeq
\vspace{-3.5em}
\begin{figure}[htbp]
  \begin{center}
    \leavevmode
\begin{picture}(0,0)%
\epsfig{file=MM.pstex}%
\end{picture}%
\setlength{\unitlength}{0.00083300in}%
\begingroup\makeatletter\ifx\SetFigFont\undefined%
\gdef\SetFigFont#1#2#3#4#5{%
\reset@font\fontsize{#1}{#2pt}%
\fontfamily{#3}\fontseries{#4}\fontshape{#5}%
\selectfont}%
\fi\endgroup%
\begin{picture}(3006,1626)(1320,-1975)
\put(3415,-845){\makebox(0,0)[lb]{\smash{\SetFigFont{10}{12.0}
{\rmdefault}{\mddefault}{\updefault}
$p\left\{\vphantom{\begin{array}{c} \\[10.5ex]\end{array}} \right.$}}}
\put(1929,-1178){\makebox(0,0)[lb]{\smash{\SetFigFont{11}{13.2}
{\rmdefault}{\mddefault}{\updefault}$\vdots$}}}
\put(1801,-1936){\makebox(0,0)[lb]{\smash{\SetFigFont{12}{14.4}
{\rmdefault}{\mddefault}{\updefault}$\M^{+\,[p]}$}}}
\put(3676,-1936){\makebox(0,0)[lb]{\smash{\SetFigFont{12}{14.4}
{\rmdefault}{\mddefault}{\updefault}$\M^{-\,[p]}$}}}
\put(3804,-953){\makebox(0,0)[lb]{\smash{\SetFigFont{11}{13.2}
{\rmdefault}{\mddefault}{\updefault}$\vdots$}}}
\put(1320,-960){\makebox(0,0)[lb]{\smash{\SetFigFont{10}{12.0}
{\rmdefault}{\mddefault}{\updefault}
$p\!+\!1\left\{\vphantom{\begin{array}{c} \\[13.5ex]\end{array}}
\right.$}}}
\end{picture}
\end{center}
\end{figure}

\noindent
Moreover, one may verify the algebraic relations
\beq
T^2=0\;,
\label{cohom}
\eeq
\beq
(S-2)\,T ~=~ 0 ~=~ T\,(S+{\cal{R}}) \;;
\label{STcom}
\eeq
i.e.~the level raising operator $T$ is nilpotent and  acts nontrivially
only between
$\M^+$ and $\M^-$:
\beq
T:\,\M^{+\,[p]} \rightarrow \M^{-\,[p+2]} \;.
\eeq
The non linear terms of the field equations \Ref{geneq} are lumped
into ${\cal C }$.  In the weak field approximation the right hand side
of this equation is negligible. Since $T$ is nilpotent, there is then
an interesting analogy between equation \Ref{geneq} and the descent
equations \cite{Zumi85}. However, these involve in general two nilpotent
operators, whereas in our case $S$ satisfies equation \Ref{Seq}
instead of being nilpotent.

\paragraph{General solution}
Let us separate  the two eigenvalues of $S $ in equation
\Ref{geneq} according to \Ref{Fpmdef}. It is easy to verify that
${\cal C}^+=0$. Thus  one gets
$$
TM^-=0,\qquad
{\cal R} M^-+T M^+={\cal C}.
$$
The first relation is automatically satisfied because of
\Ref{STcom}. In conclusion, $M^+$ is arbitrary, and
\beq
M^-~=~{\cal R}^{-1}\left({\cal C}-TM^+\right)
\label{usol}
\eeq
Thus, $M^+$ contains the independent components in $M_{\alpha\beta}$
left over by
the gauge fixing \Ref{supgec}. Comparing the Young diagrams \Ref{M+M-}
to \Ref{young} we hence recover the independent components identified
in the vector potential $A_\alpha$ after imposing the recursion
gauge.
The total and the independent number of components of $M^{[p]}$ are
respectively given by
\beq
{\> \rm dim  \> } \M^{[p]}=136 {16 \choose p},\quad
{\> \rm dim  \> } \M^{+\,[p]}=(p+1){17 \choose p+2 }.
\label{dimp}
\eeq
We give a computation of these numbers in appendix~B.  Altogether,
$M_{\alpha\beta}$ contains $983041$ independent components.  Since the
gauge fixing \Ref{supgec} is defined by covariant constraints on the
superfield, these components are necessarily expressible in terms of
representations of the supersymmetry algebra \Ref{susy}. Of course,
supersymmetry is not realized level by level. Decomposing $M^+$ into
$SO(9)$ multiplets, we find the following structure
\bea
\sum_p  {\rm dim}\,\M^{+[p]} &=& 983041 \label{983041}\\
&=& 1+(44+84+128)\times (9+16+36+126+128+231+\nonumber\\
&& \phantom{1+(44+84+128)\times (9} +432+576+594+768+924) \;.
\nonumber
\eea
The $256=(44+84+128)$ corresponds precisely to the smallest
irreducible off-shell multiplet of the 10d supersymmetry algebra
\Ref{susy} \cite{BerRoo82}. Consistently, $M^+$ forms a multiple of this
multiplet. The additional singlet in \Ref{983041} corresponds to the
fact that we have not fixed the ordinary gauge invariance.

\bigskip
\paragraph{Dual space}
For future use, let us recall that we can introduce the dual space
$\M_{\rm dual}$ of superfields by means of the bilinear form
\beq\label{sp}
\langle {F}|G\rangle=
\int d\underline \theta\; \sum_{\alpha \beta}
\widetilde{F}^{\alpha \beta}\left(\underline x, \underline
\theta\right) G_{\alpha \beta}\left(\underline x, \underline
\theta\right)
\eeq
where
$$
|G\rangle\in\M\;,\quad\langle {F}|\in\M_{\rm dual}\;,
$$
$$
\widetilde{F}^{[p]\,}{}^{\alpha\beta\,}{}_{\alpha_1\cdots\alpha_p}~=~
{1\over \left(16-p\right)! }\sum_{\alpha_{p+1},\cdots,
\alpha_{16}}\epsilon_{\alpha_1\cdots \alpha_{16} } \,
F^{[16-p]\,\alpha\beta\,,\,\alpha_{p+1}\cdots \alpha_{16}}
$$
Breaking the $O(9,1)$ invariance, one may identify $\M$ and $\M_{\rm
dual}$ by means of $\sigma^0_{\alpha\beta}$, for example. With respect
to the decomposition $SO(9,1)\rightarrow SO(2,1)\times SO(7)$, the
bilinear form \Ref{sp} then yields an $SO(7)$ invariant scalar product.
on $\M$, on which the $SO(2,1)$ generators act as
We are going to use this scalar product in the subsequent analysis of
the superfield constraints \Ref{KKK}. Note finally, that with
respect to this scalar product the operator $S$ from \Ref{Sdef} is
self-adjoint
\beq
S^{\rm ad} =S\;.
\eeq

\subsection{Extracting dynamics from the superspace constraints}

So far in this section, we have restricted the supercurl
$M_{\alpha\beta}$ only by the recursion gauge condition \Ref{supgec},
thereby restricting the gauge freedom \Ref{sugauge}. Further
restrictions and in particular dynamical equations arise from imposing
further constraints \Ref{KYM}, \Ref{KI}, and \Ref{KIR}, respectively,
on the supercurl. These constraints have been casted into the form of
projections under an operator $K$, $K^2=K$, $\overline{K}\equiv I-K$,
such that $M_{\alpha\beta}$ is subject to
\beq\label{Kdec}
\overline{K}_{\alpha \beta}^{\alpha' \beta'}M_{\alpha' \beta'}~=~0\;.
\eeq
This defines a decomposition of the superfields in $\M$ into
\beq\label{K}
\M~=~K\M+\overline{K}\M~\equiv~\M_\|+\M_\bot\;.
\eeq
The role of $K$ is twofold. First, it further restricts the field
content in the superfield $M_{\alpha\beta}$ by certain algebraic
relations; secondly, it implies field equations for the remaining
independent superfield components.

The explicit projectors for the dynamical constraints \Ref{KYM},
\Ref{KI}, and \Ref{KIR} are given by
\bea
(K_{{\rm \scriptstyle YM}})_{\alpha \beta}^{\alpha'
\beta'}&=&
{\textstyle{1\over16}}
\left(\sigma^a\right)_{\alpha
\beta}
\left(\sigma_a\right)^{\alpha' \beta'}\;,\label{KYMx}\\[3ex]
(K_{{\rm \scriptstyle I}})_{\mu \nu}^{\mu '\nu'}&=&
{\textstyle{1\over8}}\,\delta_{\mu \nu}\,\delta^{\mu '\nu'} \;,\label{KIx}\\
(K_{{\rm \scriptstyle I}})
_{\overline\mu \overline{\nu}}^{\overline\mu '\overline{\nu}'}&=&
{\textstyle{1\over8}}\,\delta_{\overline\mu \overline{\nu}}\,
\delta^{\overline\mu '\overline{\nu}'} \;,\nonumber\\
(K_{{\rm \scriptstyle I}})_{\mu \overline{\nu}}^{\mu '\overline{\nu}'}&=&
{\textstyle{1\over8}}\,\delta_{\mu \overline{\nu}}\,
\delta^{\mu '\overline{\nu}'} +
{\textstyle{1\over8}}\,\gamma^i{}_{\mu \overline{\nu}}\,
\gamma_i{}^{\mu '\overline{\nu}'} +
{\textstyle{1\over16}}\,\gamma^{ij}{}_{\mu \overline{\nu}}\,
\gamma_{ij}{}^{\mu '\overline{\nu}'}  \;,\nonumber\\[3ex]
(K_{{\rm \scriptstyle IR}})_{\mu \nu}^{\mu '\nu'}&=&0\;,
\label{KIRx}\\
(K_{{\rm \scriptstyle IR}})
_{\overline\mu \overline{\nu}}^{\overline\mu '\overline{\nu}'}&=&0
\nonumber\\
(K_{{\rm \scriptstyle IR}})_{\mu \overline{\nu}}^{\mu
'\overline{\nu}'} &=&
{\textstyle{1\over8}}\,\gamma^i{}_{\mu \overline{\nu}}\,
\gamma_i{}^{\mu '\overline{\nu}'} +
{\textstyle{1\over16}}\,\gamma^{ij}{}_{\mu \overline{\nu}}\,
\gamma_{ij}{}^{\mu '\overline{\nu}'}\;.\nonumber
\eea
as one extracts from \Ref{con10}, \Ref{dynuu} and \Ref{dynuur}
(putting for simplicity $v^i=\delta^{i8}$, cf.~appendix~A).  These
projectors are self-adjoint w.r.t.\ to the scalar product \Ref{sp},
i.e.\ $K_{{\rm \scriptstyle YM}}=K^{\rm ad}_{{\rm \scriptstyle YM}}$,
etc.

Note, that $K_{{\rm \scriptstyle I}}$ is the weakest of these
constraints in the sense that
\beq\label{weaker}
K_{{\rm \scriptstyle YM}}\M\subset K_{{\rm \scriptstyle I}}\M\;,\qquad
K_{{\rm \scriptstyle IR}}\M\subset K_{{\rm \scriptstyle I}}\M\;.
\eeq

In the following, we are going to analyze the content of these sets of
superspace constraints. To this end, we first give the general recipe
how to obtain field content and field equations implied by a
constraint of the type \Ref{Kdec} and subsequently apply this
formalism to the constraints \Ref{KYMx}, \Ref{KIx}, and \Ref{KIRx}.

\paragraph{Field content}

To identify the physical field content among the components of the
supercurl $M_{\alpha\beta}$, we collect the constraints that have been
imposed on $M_{\alpha\beta}$. These are given by the recursion gauge
expressed by \Ref{geneq} and the constraint \Ref{Kdec}:
\bea
\left(S+{\cal R}\right)\,M &=& -{{\cal R}+2\over {\cal R}}\;T\,M
+ {\cal C} \;,\label{system}\\[1ex]
\overline{K}\,M &=& 0\;. \nonumber
\eea
This obviously leaves
\beq
\M_\|^+ ~\equiv~ \M_\| \cap \M^+ ~=~
\mbox{ker}\;\overline{K} \cap \mbox{ker}\;(S+{\cal R}) \;,
\eeq
undetermined. The independent (or physical) superfield components in
$M_{\alpha\beta}$ are hence given by $\M_\|^+$, the space of
eigenvectors of the operator $KSK$ with eigenvalue $-p$. The remaining
part of $M_{\alpha\beta}$ is consequently determined by the system
\Ref{system} in terms of derivatives and nonlinear combinations of the
physical fields. The fact that this part is in fact overdetermined by
\Ref{system} then in turn implies the field equations as we shall
discuss now.

\paragraph{Field equations}

The dynamical equations arise from combining the two equations of
\Ref{system} into
\beq\label{system2}
\left(S+{\cal R}\right)M^{[p]}_\|+{{\cal R}+2\over {\cal
R}}TM^{[p-2]}_\|~=~{\cal C}^{[p]}\;.
\eeq
This defines $M^{[p]}_\|$ in terms of the lower levels unless we project
out onto vectors $\langle z|$ such that
$$
\langle z|\left(S+{\cal R}\right)M^{[p]}_\|~=~
\langle z|\left(S+{\cal R}\right)\,K\,M^{[p]}~=~0\;,
$$
in which case \Ref{system2} implies a restriction on the image of
$T$. The relevant vectors $\langle z|$ are hence simultaneous
eigenvectors of $K^{\rm ad}$ with zero eigenvalue and of $S^{\rm ad}$
with eigenvalue $2$
$$
\langle z| \left(S+{\cal R}\right) K~=~0\;,
$$
we denote them as $\langle z_\bot ^-|$.  These are eigenvectors of
$\overline{K}^{\rm ad}S^{\rm ad}\overline{K}^{\rm ad}$ with eigenvalue
$2$. (As discussed above, for the superspace constraint we find that
$S$, $K_{{\rm \scriptstyle I}}$, and $K_{{\rm \scriptstyle IR}}$ are
self-adjoint w.r.t.\ the scalar product induced by \Ref{sp}.)  For any
such eigenvector $\langle z_\bot ^-|$, \Ref{system} yields the
dynamical equation
\beq\label{dyn}
{{p\!+\!2}\over p}\left\langle z_\bot ^- \Big|
\;TM^{[p-2]}_\|\right. ~=~
\left\langle z_\bot ^-|\;{\cal C}^{[p]} \right.
\eeq
Vice versa, if \Ref{dyn} is satisfied for all vectors of the form
$\langle z_\bot ^-|$, the system \Ref{system} has a solution for
$M_{\alpha\beta}^{[p]}$ in terms of the lower levels.
\medskip

Thus, the basic information about the content of the dynamical
constraint \Ref{Kdec} concerns the set of simultaneous eigenvectors
$\langle z_\bot^-|$ and $|z_\|^+\rangle$, respectively. We denote the
corresponding spaces by $\M_\bot^-$ and $\M_\|^+$,
respectively. Counting of dimensions yields the identity
\beq\label{dimm}
{\rm dim}\, \M_\bot - {\rm dim}\, \M^+
= {\rm dim}\, \M_\bot^- - {\rm dim}\, \M_\|^+ \;,
\eeq
where the numbers on the l.h.s.~can simply be extracted from the
representation tables of $SO(9,1)$ and $SO(2,1)\times SO(7)$,
respectively. For the lowest levels, these tables are collected in
appendix~C.

\section{The physical field content.}

In this chapter, we will determine the field content which is induced
by the different dynamical constraints $K_{{\rm \scriptstyle YM}}$,
$K_{{\rm \scriptstyle I}}$, and $K_{{\rm \scriptstyle IR}}$. The
result for the latter is summarized in the tables \Ref{spectrumIR} and
\Ref{spectrumI}.  We recall, that with the strong superspace
constraint \Ref{con10}, the arbitrariness in the supercurl
$M_{\alpha\beta}$ is restricted to the levels $p=0$ and $p=1$, i.e.\
all higher levels are determined. By analyzing the Bianchi identities
for the supercurvature one verifies that in this case the following
superfield relation holds \cite{AbFoJa88}
\beq\label{recu}
{\cal R}({\cal R}\!+\!1)\,M_{\alpha\beta} ~=~
{\textstyle\frac12}\,(\sigma^{a})_{\alpha\beta} \,
(\sigma_{a}{}^{bc})_{\gamma_1\gamma_2}\,
\theta^{\gamma_1}\theta^{\gamma_2}\,F_{bc} \;,
\eeq
where
$$
F_{ab} = \partial_a A_b -\partial_b A_b + [A_a, A_b]_- \;,
$$
is now the curvature of the superfield $A_a$. Together with \Ref{AM},
one hence obtains recurrence relations which completely determine
$M_{\alpha\beta}$ in terms of its lowest components -- the physical
fields $X_a$ and $\phi^\alpha$. The field content associated with
$K_{{\rm \scriptstyle YM}}$ hence precisely coincides with the
ten-dimensional Yang-Mills multiplet.

With the integrable \Ref{KI} and the reduced integrable constraint
\Ref{KIR} the situation becomes essentially more complex. In
particular, there will be more superfield components left undetermined
by the recurrent relations, i.e.\ the spectrum of states turns out to
be considerably larger. In this section we analyze the physical field
content associated with these integrable superfield constraints
$K_{{\rm \scriptstyle IR}}$ and $K_{{\rm \scriptstyle I}}$.

According to the general discussion above, the independent components
in the supercurl $M_{\alpha\beta}$ are given by the space $\M_\|^+$,
i.e.\ by the intersection of the kernels of $(S\!+\!{\cal R})$ and
$\overline{K}$. We start from \Ref{dimm}
\beq\label{new}
{\rm dim}\, \M_\|^+ ~=~
{\rm dim}\, \M^+ - {\rm dim}\, \M_\bot + {\rm dim}\, \M_\bot^- \;,
\eeq
and will in the following determine the r.h.s.\ of this
equation for $K_{{\rm \scriptstyle IR}}$ and $K_{{\rm \scriptstyle
I}}$. To this end we first describe the decomposition of superfields
into irreducible representations of $SO(2,1)$.

\mathversion{bold}
\subsection{Decomposition into $SO(2,1)$ representations}
\mathversion{normal}

The integrable constraints \Ref{KIx}, \Ref{KIRx} are still invariant
under the action of $SO(2,1)$ corresponding to the second factor in
$SO(9,1)\rightarrow SO(2,1)\times SO(7)$. This provides a convenient
way to organize the spectrum. Explicitly, this group acts on the
supercurl as given in \Ref{so21}. The generators are pairwise
adjoint with respect to the scalar product defined in \Ref{sp}
$$
\delta_0^{\rm ad} = \delta_0\;,\quad
\delta_{\pm1} ^{\rm ad} = \delta_{\mp1}\;.
$$

The supercurl $M_{\alpha\beta}$ may hence be decomposed according to
the spin of $SO(2,1)$.  We label the total $SO(2,1)$ spin by $\ell$
and its $z$-component (i.e.\ the eigenvalue of $\delta_0$ which is
raised resp.\ lowered by $\delta_\pm $) by $\ell_0$. According to the
action of $\delta_0$, the value of $\ell_0$ is given by the difference
of barred and unbarred indices in a superfield
$M_{\alpha\beta,\gamma_1\dots\gamma_p}$.  Specifically,
$\M^{[p]}_{\ell_0}$ is spanned by vectors
\beq
M~=
\left\{\begin{array}{l}\label{vvv}
M_{\mu\nu,\mu_1\dots\mu_{k-1},\rb_1\dots\rb_{l+1}} \\[1ex]
M_{\mu\overline{\nu},\mu_1\dots\mu_{k},\rb_1\dots\rb_{l}} \\[1ex]
M_{\overline{\mu}\overline{\nu},\mu_1\dots\mu_{k+1},\rb_1\dots\rb_{l-1}}
\end{array}\right.\;,\qquad\mbox{with}\quad
p=k+l\;,\quad \ell_0={\textstyle\frac12}(l-k)\;,
\eeq
and the spin $\ell$ states are generated by highest weight
states at $\ell_0\!=\!\ell$ obtained from \Ref{vvv} by modding out the
action of $\delta_+$, i.e. satisfying
\beq\label{vvv0}
\left\{\begin{array}{rcl}
M_{\mu\nu,\mu_1\dots\mu_{k-2}[\rb_1,\rb_2\dots\rb_{l+2}]} &=& 0 \\[1ex]
kM_{\mu\overline{\nu},\mu_1\dots\mu_{k-1}[\rb_1,\rb_2\dots\rb_{l+1}]}
&=& M_{\mu\nu,\mu_1\dots\mu_{k-1},\rb_1\dots\rb_{l+1}} \\[1ex]
(k\!+\!1)\,M_{\overline{\mu}\overline{\nu},\mu_1\dots\mu_{k}
[\rb_1,\rb_2\dots\rb_{l}]}
&=&  M_{\mu\overline{\nu},\mu_1\dots\mu_{k},\rb_1\dots\rb_{l}} +
M_{\overline{\mu}\nu,\mu_1\dots\mu_{k},\rb_1\dots\rb_{l}}
\end{array}\right.
\eeq

Furthermore, the space $\M^{[p]}$ may be decomposed according to the
action of the symmetric group on the $p\!+\!2$ spinor indices. This is
most conveniently described in terms of Young diagrams, where we use
the standard notation $[a_1, \dots, a_n]$ to describe the Young
diagram with $n$ rows of length $a_1, \dots, a_n$. By $[a_1, \dots,
a_n]'$ we denote the conjugated Young diagram consisting of $n$ columns
of length $a_1, \dots, a_n$.  Each box of the Young diagrams now
represents a ${\bf 8}$ of $SO(7)$. The relations \Ref{vvv}, \Ref{vvv0}
then imply
\bea
\M^{[p]}_{\ell\not=0}
&=& [2]\times \Big([l\!-\!1,k\!+\!1]'+[l,k]'+[l\!+\!1,k\!-\!1]'\Big) +
[1,1]\times[l,k]'\;,\nonumber\\[2ex]
\M^{[p]}_{\ell=0}
&=& [2]\times [k\!+\!1,k\!-\!1]' +
[1,1]\times[k,k]'\;.
\eea
Since the decomposition \Ref{Fpmdef} of $\M^{[p]}$ has been defined
purely in terms of permuting the spinor indices, it commutes with the
action of the symmetric group. The Young diagram decomposition of the
eigenspaces $\M^+$ and $\M^-$ may be obtained from \Ref{M+M-} by
analyzing the decomposition of the Young diagrams under ${\bf
16}\mapsto{\bf 8}\!+\!{\bf 8}$.
Specifically, we find
\bea
\M^+_{\ell\not=0} &=&
[l\!+\!1,k\!+\!1]'+
[l,k\!+\!2]'+
[l,k\!+\!1,1]'+
[l\!+\!2,k]'+
[l\!+\!1,k,1]' \;,\nonumber\\[3ex]
\M^+_{\ell=0} &=&
[k\!+\!2,k]'+
[k\!+\!1,k,1]' \;.
\label{Mpkl}
\eea
In particular, this gives the dimension
$$
\mbox{dim}\;\M^{+\,[p]}_{\ell=0} ~=~ \left({8\atop k\!+\!1}\right)
                                     \left({9\atop k}\right)\,
\frac{17k+7}{(k+1)(k+2)}\;.
$$

\mathversion{bold}
\subsection{The reduced integrable constraint}
\mathversion{normal}

Here, we analyze the field content associated with the reduced
integrable constraint $K_{\scriptstyle\rm IR}$, given by \Ref{KIRx}.
This constraint may be equivalently rewritten as
\beq\label{con7rr}
K_{\scriptstyle\rm IR}M_{\mu\nu}~=~0~=~
K_{\scriptstyle\rm IR}M_{\overline{\mu}\overline{\nu}}\;,\quad
K_{\scriptstyle\rm IR}M_{\mu\overline{\nu}}~=~
\frac12\Big(M_{\mu\overline{\nu}}-M_{\nu\overline{\mu}}\Big)\;,
\eeq
and hence commutes with the action of the symmetric group on the ${\bf
8}$ spinor indices (in contrast to $K_{\scriptstyle\rm YM}$ and
$K_{\scriptstyle\rm I}$). This fact allows to completely resolve this
case without any explicit reference to the decomposition of the
superfield into irreducible representations of $SO(7)$ or $SO(9,1)$,
respectively.

According to \Ref{new} we have to determine the spaces $\M_\bot$ and
$\M^-_\bot$. We start with $\M_\bot=\overline{K}_{\scriptstyle\rm
IR}\M$. According to \Ref{con7rr} and \Ref{vvv0}, the spin $\ell$
sector of $\M_\bot$ is given by the vectors satisfying
\beq\label{Nspin}
\left\{\begin{array}{rcl}
M_{\mu\nu,\mu_1\dots\mu_{k-2}[\rb_1,\rb_2\dots\rb_{l+2}]} &=& 0 \\[1ex]
kM_{\mu\overline{\nu},\mu_1\dots\mu_{k-1}[\rb_1,\rb_2\dots\rb_{l+1}]}
&=& M_{\mu\nu,\mu_1\dots\mu_{k-1},\rb_1\dots\rb_{l+1}} \\[1ex]
(k\!+\!1)\,M_{\overline{\mu}\overline{\nu},\mu_1\dots\mu_{k}
[\rb_1,\rb_2\dots\rb_{l}]}
&=&  2M_{\mu\overline{\nu},\mu_1\dots\mu_{k},\rb_1\dots\rb_{l}}
\end{array}\right.
\eeq
In other words, each vector is given by its part
$M_{\overline{\mu}\overline{\nu}}$, satisfying the constraint
$$
M_{\overline{\mu}\overline{\nu},\mu_1\dots\mu_{k-2}
[\rb_1\rb_2\rb_3,\rb_4\dots\rb_{l+2}]}~=~0 \;,
$$
the other parts of $M$ are determined from this by
\Ref{Nspin}. This gives the Young diagram decomposition
of $\overline{K}_{\scriptstyle\rm IR}\,\M$:
\bea\label{Mbkl}
\overline{K}_{\scriptstyle\rm IR}\,\M_{\ell\not=0}
&=&{}2\cdot[l\!+\!1,k,1]' +2\cdot[l,k\!+\!1,1]' +[l\!+\!2,k]'
+ [l,k\!+\!2]'\\[1ex]
&&{}+[l\!-\!1,k\!+\!1,1,1]' + [l\!+\!1,k\!-\!1,1,1]'
+ [l\!+\!2,k\!-\!1,1]' \nonumber\\[1ex]
&&{}+ [l\!-\!1,k\!+\!2,1]' +[l\!+\!1,k\!+\!1]'+ [l,k,1,1]'  \;,
\nonumber\\[3ex]
\overline{K}_{\scriptstyle\rm IR}\,\M_{\ell=0}
&=& [k\!+\!1,k,1]'+[k\!+\!2,k]'
+[k\!+\!1,k\!-\!1,1,1]'+[k\!+\!2,k\!-\!1,2]' \;.\nonumber
\eea
\medskip

It remains to determine $\M^-_\bot$, the space of common eigenvectors
of $S$ and $\overline{K}_{\scriptstyle\rm IR}$. For this, we consider
the operator $S_{\scriptstyle\rm IR}\equiv\overline{K}_{\scriptstyle\rm
IR}S\overline{K}_{\scriptstyle\rm IR}$ whose action on
$M_{\overline{\mu}\overline{\nu}}$ is found from \Ref{Sdef} and
\Ref{Nspin} to be:
\bea\label{actionS}
(S_{\scriptstyle\rm IR}\,M^{[p]})
_{\overline{\mu}\overline{\nu},
\mu_1\dots\mu_{k+1},\rb_1\dots\rb_{l-1}}
&=&
2(l\!-\!1)\,M^{[p]}_{\overline{\mu}\rb_1,\mu_1\dots\mu_{k+1},
\overline{\nu}\rb_2\dots\rb_{l-1}}
\\[1ex]
&&{}-(k\!+\!1)\,M^{[p]}_{\overline{\mu}\mu_1,
\overline{\nu}\mu_2\dots\mu_{k+1},\rb_1\dots\rb_{l-1}}
\nonumber\\[1ex]
&&{}+(k\!+\!1)(l\!-\!1)\,
M^{[p]}_{\overline{\mu}\mu_1,\rb_1\mu_2\dots\mu_{k+1},
\overline{\nu}\rb_2\dots\rb_{l-1}}\;.\nonumber
\eea
The operator $S_{\scriptstyle\rm IR}$ obviously does not commute with
its ancestor $S$ and correspondingly has eigenvalues which do not
necessarily coincide with those of $S$. However, since
$\overline{K}_{\scriptstyle\rm IR}$ is an orthogonal projector, it
follows that the eigenvalues of $S_{\scriptstyle\rm IR}$ lie in the
interval $[-p,2]$. Moreover, eigenvectors of $S_{\scriptstyle\rm IR}$
with eigenvalues $-p$ and $2$, respectively, are necessarily also
eigenvectors of $S$.  Diagonalizing the action \Ref{actionS},
$S_{\scriptstyle\rm IR}$ finally may be decomposed into projectors
${\cal P}_{[\dots]}$ onto the Young diagrams from \Ref{Mbkl},
respectively:
\bea\label{SRkl}
S_{\scriptstyle\rm IR} &=&
2\Big(
{\cal{P}}_{[l-1,k+1,1,1]}+
{\cal{P}}_{[l+1,k-1,1,1]}+
{\cal{P}}_{[l,k,1,1]}+
{\cal{P}}_{[l+2,k-1,1]}\Big)\\[0.5ex]
&&{}+2\Big(
{\cal{P}}_{[l-1,k+2,1]}+
{\cal{P}}_{[l+1,k,1]}+
{\cal{P}}_{[l,k+1,1]}\Big)
+(2\!-\!k)\,{\cal{P}}_{[l+2,k]}\nonumber\\[1ex]
&&{}+
(1\!-\!l)\,{\cal{P}}_{[l,k+2]}-
{\textstyle\frac{k+2l}{2}}\,{\cal{P}}_{[l,k+1,1]}+
{\textstyle\frac{2-k-l}{2}}\,{\cal{P}}_{[l+1,k+1]}-
{\textstyle\frac{2k+l-1}{2}}\,{\cal{P}}_{[l+1,k,1]}\;,
\nonumber\\[1ex]
&&{}\mbox{for $l>k$}\;,\nonumber\\[2ex]
S_{\scriptstyle\rm IR} &=&
2\Big(
{\cal{P}}_{[k+1,k-1,1,1]}+
{\cal{P}}_{[k+2,k-1,1]}\Big)+(2\!-\!k)\,{\cal{P}}_{[k+2,k]}+
\textstyle{\frac{1-3k}2}\,
{\cal{P}}_{[k+1,k,1]}\;,\nonumber\\[1ex]
&&{}\mbox{for $l=k$}\;.\nonumber
\eea
The eigenspaces with eigenvalue 2 in this decomposition
span the space $\M_\bot^-$. Putting \Ref{Mpkl}, \Ref{Mbkl}, and
\Ref{SRkl} together, we find from \Ref{new}
\beq\label{specIR}
{\rm dim} \M^+_\| ~=~
\left\{\begin{array}{cl}
{\rm
dim}([l\!+\!2]')=\left(8\atop{l\!+\!2}\right)
\quad&\mbox{for}\;\; k=0 \\[3ex]
0\quad&\mbox{otherwise} \end{array}\right.\;\;\;.
\eeq
The exceptional role of $k\!=\!0$ stems from the fact that for this
value the eigenvalue of the corresponding Young diagram $[l\!+\!2,k]'$
in \Ref{SRkl} takes the extremal values 2 such that at $k\!=\!0$ this
eigenspace becomes part of $\M_\bot^-$.

With \Ref{new} we hence have obtained the entire physical field
content in the superfield $M_{\alpha\beta}$ induced by the reduced
integrable superspace constraint $K_{\scriptstyle\rm IR}$. We collect
the result in the following table, organized by level $p$ and
$SO(2,1)$ spin $\ell$:
\beq\label{spectrumIR}
\mbox{
\begin{tabular}{|c||c|c|c|c|c|c|c|} \hline
$\ell$ &0&$\frac12$&1&$\frac32$&2&$\frac52$&3 \\[0.1em]
\hline\hline
$p\!=\!0$&{\bf 7}+{\bf 21}&&&&&&\\[0.1em]\hline
$p\!=\!1$&&{\bf 8}+{\bf 48}&&&&& \\[0.1em] \hline
$p\!=\!2$&&&{\bf 1}+{\bf 7}+{\bf 27}+{\bf 35}
&&&&\\[0.1em] \hline
$p\!=\!3$&&&&{\bf 8}+{\bf 48}&&&\\[0.1em] \hline
$p\!=\!4$&&&&&{\bf 7}+{\bf 21}&&\\[0.1em] \hline
$p\!=\!5$&&&&&&{\bf 8}&\\[0.1em] \hline
$p\!=\!6$&&&&&&&{\bf 1}\\[0.1em] \hline
\end{tabular}}
\eeq
\medskip

\noindent
The total number of states is $769=384+384+1$, where $384+384$
corresponds to 3 copies of the irreducible off-shell multiplet
$(128\!+\!128)$ of \cite{BerRoo82} and the singlet captures the remaining
bosonic gauge freedom. This counting is the first hint, that the field
content of the reduced integrable constraint remains completely
off-shell, a fact that we shall show in the next chapters.

\bigskip

\subsection{The integrable constraint}

Here, we analyze the field content associated with the integrable
constraint $K_{\scriptstyle\rm I}$, given by \Ref{KIx}. Comparing to
the result \Ref{spectrumIR} for the stronger constraint
$K_{\scriptstyle\rm IR}$, even more fields must appear in this
case. Note that in this section $\M_\|$ and $\M_\bot$ refer to the
decomposition \Ref{K} with respect to $K_{\scriptstyle\rm
I}$. Nevertheless, it is $\overline{K}_{\scriptstyle\rm
I}\,\M=\overline{K}_{\scriptstyle\rm IR}\,\M$.
To make use of the result of the previous section, we rewrite
\Ref{new} as
\bea
{\rm dim}\, \M_\|^+ &=&
{\rm dim}\, \M^+ - {\rm dim}\, {\cal M}_\bot
+ {\rm dim}\, {\cal M}_\bot^- \label{neww}\\[1ex]
&=& \Big( {\rm dim}\,\M^+ -
{\rm dim}\,({\overline{K}_{\scriptstyle\rm IR}\M})^+\Big) +
{\rm dim}\;({\overline{K}_{\scriptstyle\rm IR}\M})^+_\|  \;.
\nonumber
\eea
where the term in the brackets on the r.h.s. has been determined in
\Ref{specIR} above and $({\overline{K}_{\scriptstyle\rm IR}\M})^+_\|$
is defined to be the intersection of
\bea
({\overline{K}_{\scriptstyle\rm IR}\M})^+ &\equiv&
(S_{\scriptstyle\rm IR}\!-\!2)\,{\overline{K}_{\scriptstyle\rm IR}\M}
\;,\nonumber\\[1ex]
\mbox{and}\quad
({\overline{K}_{\scriptstyle\rm IR}\M})_\| &\equiv&
{\overline{K}_{\scriptstyle\rm IR}K_{\scriptstyle\rm I}\M}  \;.
\nonumber
\eea
This space hence contains the fields that enlarge the spectrum with
respect to \Ref{spectrumIR}. Its dimension remains to be computed.

We first consider the case $l\!=\!k$, i.e.\ the $SO(2,1)$ singlets.
The space $({\overline{K}_{\scriptstyle\rm IR}\M})^+_\|$ then is
generated by vectors $v=v_1+v_2$ such that $v_1$ and $v_2$ are
eigenvectors of $S_{\scriptstyle\rm IR}$ with eigenvalues $(2\!-\!k)$
and ${\textstyle \frac12}\,(1\!-\!3k)$, respectively -- cf. \Ref{SRkl}
--, which in particular satisfy
\beq\label{tech}
K_{\scriptstyle\rm I}S_{\scriptstyle\rm IR}\,v~=~
K_{\scriptstyle\rm I}S_{\scriptstyle\rm IR}K_{\scriptstyle\rm I}
\;v~=~{\textstyle \frac14}\,(k\!-\!2)\,v\;.
\eeq
The second equality is obtained from contracting \Ref{actionS} over
$\overline{\mu\nu}$. Since $K_{\scriptstyle\rm I}$ is an orthogonal
projector, comparing \Ref{tech} to the eigenvalues of
$S_{\scriptstyle\rm IR}$ shows, that it can be satisfied only if
\beq
{\textstyle \frac12}\,(1\!-\!3k)~\le~
{\textstyle \frac14}\,(k\!-\!2)~\le~ (2\!-\!k)
\qquad
\Longrightarrow
\qquad p=2k~\le~4\;.
\eeq

For $p\!=\!4$ one may show by a similar but slightly
more complicated analysis of the operator $(K_{\scriptstyle\rm
I}S_{\scriptstyle\rm IR}\,K_{\scriptstyle\rm I}S_{\scriptstyle\rm
IR}\,K_{\scriptstyle\rm I})$, that $({\overline{K}_{\scriptstyle\rm
IR}\M})^+_\|$ is empty also at this level. We leave details to the
reader.  At $p\!=\!2$, in contrast, it is
${\overline{K}_{\scriptstyle\rm IR}\M}
=({\overline{K}_{\scriptstyle\rm IR}\M})^+$ and hence ${\rm
dim}\,({\overline{K}_{\scriptstyle\rm IR}\M})^+_\|={\rm
dim}\,({\overline{K}_{\scriptstyle\rm IR}K_{\scriptstyle\rm
I}\M})=7\!+\!21$.

For states of arbitrary $SO(2,1)$ spin $\ell=\frac12(l\!-\!k)$ one
shows by similar reasoning, that ${\rm
dim}\,({\overline{K}_{\scriptstyle\rm IR}\M})^+_\| =
\left(8\atop{k+1}\right)$ iff $l\!=\!1$. The complete result is
summarized in the following table:
\beq\label{spectrumI}
\mbox{
\begin{tabular}{|l||c|c|c|c|c|c|c|} \hline
$\ell$ &0&$\frac12$&1&$\frac32$&2&$\frac52$&3 \\[0.1em]
\hline\hline
$p\!=\!0$&{\bf 7}+{\bf 21}&&{\bf 1}&&&&\\[0.1em]\hline
$p\!=\!1$&&{\bf 8}+{\bf 8}+{\bf 48}&&&&& \\[0.1em] \hline
$p\!=\!2$&{\bf 7}+{\bf 21}&&{\bf 1}+{\bf 7}+{\bf 27}+{\bf 35}
&&&&\\[0.1em] \hline
$p\!=\!3$&&&&{\bf 8}+{\bf 48}&&&\\[0.1em] \hline
$p\!=\!4$&&&&&{\bf 7}+{\bf 21}&&\\[0.1em] \hline
$p\!=\!5$&&&&&&{\bf 8}&\\[0.1em] \hline
$p\!=\!6$&&&&&&&{\bf 1}\\[0.1em] \hline
\end{tabular}}
\eeq
\medskip

\noindent
The total number of fields in this case is $816=416+400$.

\section{Recurrent relations.}

Having determined the field content, we will derive the recurrent
relations which explicitly determine the higher level superfield
components in terms of the lower level components. For simplicity, we
restrict for the rest of the paper to purely bosonic configurations,
e.g.~we set all fermionic fields to zero. This is just for the sake of
clarity, the techniques may likewise be applied to determine the
structure of the fermionic fields and field equations. In particular,
supersymmetry is unbroken up to this point.

For the strong superfield constraint, the superfield $M_{\alpha\beta}$
is entirely determined by its zero level components which are the
physical Yang-Mills fields. The recurrent relations which determine
the higher superfield levels have been given in \Ref{recu}.  With the
integrable constraints \Ref{KI}, \Ref{KIR} the picture becomes
more complicated. In view of the field content \Ref{spectrumIR} and
\Ref{spectrumI}, a huge amount of additional fields has to be
introduced to eventually obtain closed recurrence relations which
replace \Ref{recu}.

To keep things tractable, we will most of the time restrict the
analysis to those fields which transform as singlets under the
$SO(2,1)$ symmetry underlying \Ref{BZ}. Despite this technical
restriction, the method described in the following allows
straight-forward although more tedious generalization to the higher
spin fields.

\subsection{The reduced integrable constraint}

In the spin zero sector the field content associated with $K_{{\rm
\scriptstyle IR}}$ according to \Ref{spectrumIR} contains an
antisymmetric tensor field in addition to the seven dimensional
Yang-Mills vector field. We have shown that in principle all higher
levels $p>0$ of $M_{\alpha\beta}$ are uniquely determined in terms of
these fields.

To make this dependence explicit, we start from the system
\Ref{system}
\beq\label{systemIR}
(S+p)\,M^{[p]} ~=~ -{\textstyle\frac{p+2}{p}}\,TM^{[p-2]}
+{\cal C}^{[p]}\;.
\eeq
Contracting this equation with $K_{{\rm\scriptstyle IR}}$, we obtain
\beq\label{systemn}
(K_{{\rm\scriptstyle IR}}S{K_{{\rm\scriptstyle IR}}}
+pK_{{\rm\scriptstyle IR}})\,M^{[p]} ~=~
-{\textstyle\frac{p+2}{p}}\,K_{{\rm\scriptstyle IR}}TM^{[p-2]}
+K_{{\rm\scriptstyle IR}}{\cal C}^{[p]}\;.
\eeq
Recall, that the space $\M_{\ell=0}$ decomposes into
\bea
\overline{K}_{\scriptstyle\rm IR}\,\M
&=& [k\!+\!1,k,1]'+[k\!+\!2,k]'
+[k\!+\!1,k\!-\!1,1,1]'+[k\!+\!2,k\!-\!1,2]' \nonumber\\[2ex]
K_{\scriptstyle\rm IR}\,\M&=& [k\!+\!1,k,1]'+[k\!+\!2,k]'+[k,k,2]'
\nonumber
\eea
Together with the eigenvalue decomposition \Ref{Mpkl} of $S$ this
gives the eigenvalues of the operator $(K_{{\rm\scriptstyle
IR}}S{K_{{\rm\scriptstyle IR}}})$ at each level $p$, which are $2$,
$-\frac{p}2$, and $\frac14(6-p)$. This allows to effectively invert
the system \Ref{systemn} to obtain
\bea
M^{[p]}&=&\Xi^{[p]}\,TM^{[p-2]}
-{\textstyle\frac{p}{p+2}}\,\Xi^{[p]}\,{\cal C}^{[p]} \;,
\label{recuIR}\\[3ex]
\mbox{with}\quad\Xi^{[p]}&=&\left\{ -{\textstyle\frac8{3p^2(p+2)}}\,
(K_{{\rm\scriptstyle IR}}S{K_{{\rm\scriptstyle IR}}})^2
+{\textstyle\frac{2(p+14)}{3p^2(p+2)}}\,
(K_{{\rm\scriptstyle IR}}S{K_{{\rm\scriptstyle IR}}})
-{\textstyle\frac{3p^2+10p+24}{3p^2(p+2)}}\,
K_{{\rm\scriptstyle IR}} \right\} \;.
\nonumber
\eea
Thus, we have completely resolved this sector with an explicit
recurrent definition of all higher level components of the superfield
$M_{\alpha\beta}$. In a similar way, one may obtain the defining
relations for the higher $SO(2,1)$ spin sectors.

As an illustration, we evaluate the lowest two bosonic levels of the
supercurl $M_{\alpha\beta}$. They are determined by the level zero
fields: the seven dimensional Yang-Mills vector field $X_i$ and an
antisymmetric tensor field $B_{ij}$:
\bea
M^{[0]}_{\mu \overline{\nu}} &=&
\gamma^i{}_{\mu \overline{\nu}} \, X_i +
\gamma^{ij}{}_{\mu \overline{\nu}} \, B_{ij} \;,\\[4ex]
M^{[2]}_{\mu \overline{\nu}} &=&
\left(\textstyle{\frac12}\,
\gamma^{ij}_{\mu \overline{\nu}}\,\delta_{\rho_1\rb_2}
-\textstyle{\frac12}\,
\gamma^{m}_{\mu \overline{\nu}}\,\gamma^{ijm}_{\rho_1\rb_2}
+\textstyle{\frac14}\gamma^{mn}_{\mu \overline{\nu}}
\,\gamma^{ijmn}_{\rho_1\rb_2} \right)\,
\theta^{\rho_1} \theta^{\rb_2} \;Y_{ij}
\nonumber\\[1ex]
&&{}+\left(2\,\gamma^{i}_{\mu \overline{\nu}}\,\delta_{\rho_1\rb_2}
+\gamma^{mn}_{\mu \overline{\nu}}\,\gamma^{imn}_{\rho_1\rb_2}\right)\,
\theta^{\rho_1} \theta^{\rb_2} \;{\cal D}^jB_{ij}
\nonumber\\[1ex]
&&{}+\left(
\gamma^{m}_{\mu \overline{\nu}}\,\gamma^{ijkm}_{\rho_1\rb_2} -
\gamma^{km}_{\mu \overline{\nu}}\,\gamma^{ijm}_{\rho_1\rb_2} +
2\,\gamma^{km}_{\mu \overline{\nu}}\,\gamma^{ijm}_{\rho_1\rb_2}\right)\,
\theta^{\rho_1} \theta^{\rb_2} \;{\cal D}_k\,B_{ij}
\nonumber\\[1ex]
&&{} +
\left(2\,\gamma^{m}_{\mu \overline{\nu}}\gamma^{ijm}_{\rho_1\rb_2}
+2\,\gamma^{ij}_{\mu \overline{\nu}}\delta_{\rho_1\rb_2}
+\gamma^{mn}_{\mu \overline{\nu}}\gamma^{ijmn}_{\rho_1\rb_2}\right)\,
\theta^{\rho_1} \theta^{\rb_2} \; \left[B_{ik},B_{j}{}^{k}\right]
\nonumber\\[1ex]
&&{}+2\left(\gamma^{l}_{\mu \overline{\nu}}\gamma^{ijk}_{\rho_1\rb_2}
+\gamma^{lm}_{\mu \overline{\nu}}\gamma^{ijkm}_{\rho_1\rb_2} \right)\,
\theta^{\rho_1} \theta^{\rb_2} \; \left[B_{ij},B_{kl}\right] \;.
\nonumber
\eea
Here, $Y_{ij}$ is the Yang-Mills field strength
\beq
Y_{ij}=\partial_iX_j-\partial_jX_i +\left[X_i,\, X_j\right]_-\;,
\label{Ydef}
\eeq
and ${\cal D}_k$ denotes the gauge covariant derivative
\beq
{\cal D}_k\,B_{ij} ~=~ \partial_k\,B_{ij}+[X_k,B_{ij}] \;.
\eeq

\subsection{The integrable constraint}

For the reduced integrable $K_{{\rm \scriptstyle IR}}$ constraint, we
have given the complete recurrent solution \Ref{recuIR}. Since the
integrable constraint $K_{{\rm \scriptstyle I}}$ is weaker in the
sense of \Ref{weaker}, its field content is larger and the recurrent
relations will involve more fields. From \Ref{spectrumI} we know
already that the spectrum associated with $K_{{\rm \scriptstyle I}}$
in its spin 0 sector contains another copy of the vector and tensor
fields.

The system \Ref{systemIR} in this case gives rise to the recurrent
relations
\bea\label{recuI}
M^{[p]}&=&\Xi^{[p]}\,\left(TM^{[p-2]}
-{\textstyle\frac{p}{p+2}}\,{\cal C}^{[p]}\right)
+ \left\{{\textstyle\frac{p}{p+2}}\,\Xi^{[p]}
S+ I \right\}
K_{{\rm \scriptstyle I}}\overline{K}_{{\rm \scriptstyle IR}}
N^{[p]}\;,
\eea
where $\Xi^{[p]}$ has been defined in \Ref{recuIR} above, and $N$ is a
superfield which satisfies:
\bea
N&=&K_{{\rm \scriptstyle I}}\overline{K}_{{\rm \scriptstyle IR}}\,N
\quad\Longrightarrow\quad
N=\left\{\begin{array}{l} N_{\mu\nu} = \delta_{\mu\nu}\;
\theta^{\rb}\partial_\rho n\\
N_{\mu\overline{\nu}} = \delta_{\mu\overline{\nu}}\,n\\
N_{\overline{\mu\nu}} =
\delta_{\overline{\mu\nu}}\;
\theta^{\rho}\partial_{\rb}n\end{array}\right.\;\;\,
\eea
with a scalar superfield $n$, further constrained by
\Ref{vvv0}. Taking different projections of \Ref{systemIR} one may
obtain the remaining recurrent relations which determine the higher
levels of these scalar superfields in terms of the lower levels in
$\M_{\alpha\beta}$ and $n$. These however becomes more
tedious due to the fact that the integrable constraint can no longer be
expressed entirely in terms of permuting spinor indices.

Here, we restrict to giving the first two levels of $n$
which have the particularly simple form
\bea
n{}\!^{[0]}&=& 0\\[2ex]
n{}\!^{[2]}&=&
\gamma^i{}_{\rho_1 \rb_2} \, \theta^{\rho_1}\theta^{\rb_2}\,Z_i +
\gamma^{ij}{}_{\rho_1 \rb_2} \, \theta^{\rho_1}\theta^{\rb_2}\,C_{ij} \;.
\nonumber
\eea
The level $p=2$ is completely undetermined and hence contains the
additional physical fields denoted by $Z_i$ and $C_{ij}$ whose
existence has been anticipated in \Ref{spectrumI}. Evaluating
\Ref{recuI} we find for the supercurl\footnote{With respect to a
previous version of this article, the component fields have been
redefined in order to make the structure more transparent. In
particular, this brings the field equations into a much simpler form.}
\bea
M^{[0]}_{\mu \overline{\nu}} &=&
\gamma^i{}_{\mu \overline{\nu}} \, X_i +
\gamma^{ij}{}_{\mu \overline{\nu}} \, B_{ij} \;,\label{MMMI}\\[4ex]
M^{[2]}_{\mu \nu} &=& \delta_{\mu \nu}\,\gamma^{i}_{\rb_1\rb_2} \,
\theta^{\rb_1} \theta^{\rb_2}\;Z_i
+\delta_{\mu \nu}\,\gamma^{ij}_{\rb_1\rb_2}\,
\theta^{\rb_1} \theta^{\rb_2}\;C_{ij} \;,
\nonumber\\[3ex]
M^{[2]}_{\mu \overline{\nu}} &=&
\left(
\delta_{\mu \overline{\nu}}\,\gamma^{i}_{\rho_1\rb_2}
+{\textstyle{\frac12}}\,
\gamma^{i}_{\mu \overline{\nu}}\,\delta_{\rho_1\rb_2}
+ {\textstyle{\frac14}}\,
\gamma^{mn}_{\mu \overline{\nu}}\,\gamma^{imn}_{\rho_1\rb_2} \right)\,
\theta^{\rho_1} \theta^{\rb_2} \; Z_i
\nonumber\\[1ex]
&&{}+\left(
\delta_{\mu \overline{\nu}}\,\gamma^{ij}_{\rho_1\rb_2}
+{\textstyle{\frac12}}
\,\gamma^{ij}_{\mu \overline{\nu}}\,\delta_{\rho_1\rb_2}
+{\textstyle{\frac12}}
\,\gamma^{m}_{\mu \overline{\nu}}\,\gamma^{ijm}_{\rho_1\rb_2}
+{\textstyle{\frac14}}
\,\gamma^{mn}_{\mu \overline{\nu}}\,\gamma^{ijmn}_{\rho_1\rb_2} \right)\,
\theta^{\rho_1} \theta^{\rb_2} \; C_{ij}
\nonumber\\[2ex]
&&{}+
\left(\textstyle{\frac12}\,
\gamma^{ij}_{\mu \overline{\nu}}\,\delta_{\rho_1\rb_2}
-\textstyle{\frac12}\,
\gamma^{m}_{\mu \overline{\nu}}\,\gamma^{ijm}_{\rho_1\rb_2}
+\textstyle{\frac14}\gamma^{mn}_{\mu \overline{\nu}}
\,\gamma^{ijmn}_{\rho_1\rb_2} \right)\,
\theta^{\rho_1} \theta^{\rb_2} \;Y_{ij}
\nonumber\\[1ex]
&&{}+\left(2\,\gamma^{i}_{\mu \overline{\nu}}\,\delta_{\rho_1\rb_2}
+\gamma^{mn}_{\mu \overline{\nu}}\,\gamma^{imn}_{\rho_1\rb_2}\right)\,
\theta^{\rho_1} \theta^{\rb_2} \;{\cal D}^jB_{ij}
\nonumber\\[1ex]
&&{}+\left(
\gamma^{m}_{\mu \overline{\nu}}\,\gamma^{ijkm}_{\rho_1\rb_2} -
\gamma^{km}_{\mu \overline{\nu}}\,\gamma^{ijm}_{\rho_1\rb_2} +
2\,\gamma^{km}_{\mu \overline{\nu}}\,\gamma^{ijm}_{\rho_1\rb_2}\right)\,
\theta^{\rho_1} \theta^{\rb_2} \;{\cal D}_k\,B_{ij}
\nonumber\\[1ex]
&&{} +
\left(2\,\gamma^{m}_{\mu \overline{\nu}}\gamma^{ijm}_{\rho_1\rb_2}
+2\,\gamma^{ij}_{\mu \overline{\nu}}\delta_{\rho_1\rb_2}
+\gamma^{mn}_{\mu \overline{\nu}}\gamma^{ijmn}_{\rho_1\rb_2}\right)\,
\theta^{\rho_1} \theta^{\rb_2} \; \left[B_{ik},B_{j}{}^{k}\right]
\nonumber\\[1ex]
&&{}+2\left(\gamma^{l}_{\mu \overline{\nu}}\gamma^{ijk}_{\rho_1\rb_2}
+\gamma^{lm}_{\mu \overline{\nu}}\gamma^{ijkm}_{\rho_1\rb_2} \right)\,
\theta^{\rho_1} \theta^{\rb_2} \; \left[B_{ij},B_{kl}\right] \;,
\nonumber\\[3ex]
M^{[2]}_{\overline\mu \overline{\nu}} &=&
\delta_{\overline\mu \overline{\nu}}\,\gamma^{i}_{\rho_1\rho_2} \,
\theta^{\rho_1} \theta^{\rho_2}\;Z_i
+\delta_{\overline\mu \overline{\nu}}\,\gamma^{ij}_{\rho_1\rho_2}\,
\theta^{\rho_1} \theta^{\rho_2}\;C_{ij} \;.{}
\nonumber
\eea

Summarizing, we have shown that in the sector of $SO(2,1)$ singlets,
the supercurl $M_{\alpha\beta}$ in recursion gauge and with the
integrable superspace constraint \Ref{KIx} imposed, is determined in
all orders by the set of physical fields
\beq\label{nef}
X_i\,,\;B_{ij}\,,\;Z_i\,,\;C_{ij} \;,
\eeq
which enter as components at the levels $p\!=\!0$ and $p\!=\!2$ of the
superfield expansion of $M_{\alpha\beta}$ as given in \Ref{MMMI}. In
the following, we will study what kind of dynamical relations we may
in addition extract for these fields.

Since $K_{\scriptstyle\rm I}$ is the weakest of the three constraints
we are studying, the other two cases may be embedded as particular
truncations of \Ref{MMMI}. It is easy to see that they correspond to
\bea
K_{\scriptstyle\rm YM} &:& C_{ij}=Y_{ij}\,,\;Z_i=0\,,\;B_{ij}=0\;,
\label{emb}\\[1ex]
K_{\scriptstyle\rm IR} &:& C_{ij}=0\,,\;Z_i=0\;.\nonumber
\eea

\section{Field equations.}

In this section we will determine the field equations implied by the
integrable constraints $K_{\scriptstyle\rm I}$, $K_{\scriptstyle\rm
IR}$ for the physical fields. As in the previous chapter we restrict
to the dimensionally reduced situation where all fields depend only on
the coordinates $x^i$, $i=1, \dots, 7$, thereby consistently
truncating the system to singlets under $SO(2,1)$.

The field content in this sector is given by \Ref{nef} for
$K_{\scriptstyle\rm I}$, $K_{\scriptstyle\rm IR}$ implies the further
truncation \Ref{emb}. Following the general discussion of section 3,
the dynamical content arises from projecting the image of the operator
$T$ according to \Ref{dyn} onto the space $\M_\bot^-$. Applying this
to the integrable constraints, we find that $K_{\scriptstyle\rm IR}$
in fact does not imply any field equation on the fields $X_i$,
$B_{ij}$, such that the corresponding spectrum remains completely
off-shell. The weaker constraint $K_{\scriptstyle\rm I}$ which has a
larger spectrum, will give rise to first order dynamical equations for
the additional fields $Z_i$ and $C_{ij}$, coupled to the off-shell
fields $X_i$ and $B_{ij}$.

Let us first recapitulate the case of the strong constraint. At level
$p\!=\!2$ the supercurl is uniquely determined by the Yang-Mills
fields according to the
lowest order component of equation \Ref{recu}
\beq\label{rec2}
M^{[2]}_{\alpha\beta} =
(\sigma^{a})_{\alpha\beta} \,
(\sigma_{a}{}^{bc})_{\gamma_1\gamma_2}\,
\theta^{\gamma_1}\theta^{\gamma_2}\,Y_{bc} \;.
\eeq
Since there are no new fields arising on this level, relation
\Ref{dimm} yields
\beq\label{dimn}
\M_\|^+= \emptyset \quad\Longrightarrow\quad
{\rm dim}\, \M_\bot^-~=~ {\rm dim}\, \M_\bot - {\rm dim}\, \M^+ \;,
\eeq
where the numbers on the r.h.s.\ may be extracted from the
representation tables collected in appendix~C.  In particular, this
shows that no field equations arise on this level.

At level $p\!=\!4$, equation \Ref{dimn} with table \Ref{Tp4so10} shows
that $\M^-_\bot$ is nonempty but contains e.g.\ the vector
representation ${\bf 10}$ of $SO(9,1)$. According to \Ref{dyn} the
dynamical equation is given by the scalar product
\beq\label{sc10}
\left\langle \M_\bot^-\,{}^{[4]}_{\bf 10}{}\,\right| \;
\left(TM^{[2]}_\| - {\textstyle{2\over 3}} {\cal C}^{[4]} \right) \;,
\qquad\mbox{with}\quad
TM^{[2]}_\| \sim \partial^b\,Y_{ab}
\eeq

Since this is the nondegenerate scalar product on a space of
multiplicity one, it suffices to show that $TM^{[2]}_\|\not=0$ (with
$M^{[2]}_\|$ given by \Ref{rec2}) to indeed obtain the bosonic part of
the Yang-Mills field equations
\beq\label{YMc}
{\cal D}^b\,Y_{ab} ~=~ 0 \;.
\eeq
One might expect to find further relations in the $SO(9,1)$
representations ${\bf 120}$ and ${\bf 126}$, respectively, in which
according to \Ref{Tp4so10} the space $\M^-_\bot$ also has nonvanishing
contributions. However, the first one contains precisely the Bianchi
identities of $Y_{ab}$ which are automatically satisfied, whereas
there is no nontrivial image of $T$ into the ${\bf 126}$ as one may
easily verify. Thus, in agreement with \cite{AbFoJa88}, there arise no
further restrictions than the Yang-Mills equations of motion
\Ref{YMc}, here.

In the following, we repeat this analysis for the integrable
constraints $K_{\scriptstyle\rm I}$ and $K_{\scriptstyle\rm IR}$.

\mathversion{bold}
\subsection{The reduced integrable constraint}
\mathversion{normal}

For the reduced integrable constraint $K_{\scriptstyle\rm IR}$, the
entire constraint is encoded in the system \Ref{systemIR} of which we
have already solved the projection \Ref{systemn} by imposing the
recurrent relations \Ref{recuIR}. It remains to study the
complementary projection:
\beq
\overline{K}_{\scriptstyle\rm IR}\,(S+p)\,M^{[p]}~=~
-{\textstyle\frac{p+2}{p}}\,\overline{K}_{\scriptstyle\rm IR}\,
TM^{[p-2]}
+\overline{K}_{\scriptstyle\rm IR}\,{\cal C}^{[p]}
\;.
\eeq
Plugging in the explicit solution \Ref{recuIR}, one obtains after some
calculation
\beq\label{dynIR}
\overline{K}_{\scriptstyle\rm IR}
\left(2S+(p\!-\!4)\right)\overline{K}_{\scriptstyle\rm IR}
\left(4S+(3p\!-\!2)\right)\overline{K}_{\scriptstyle\rm IR}\;
\left(TM^{[p-2]}-{\textstyle\frac{p}{p+2}}\,{\cal C}^{[p]}
\right) ~=~0\;.
\eeq

This encodes the entire dynamics of this constraint. Comparing
\Ref{dynIR} with \Ref{SRkl}, one recognizes \Ref{dyn}; the operator on
the l.h.s.\ of this equation is precisely the projector onto
$\M_\bot^-$.

However, it turns out (as we have explicitly checked for $p\le6$ and
are confident that it holds on all levels) that equation \Ref{dynIR}
becomes an identity when $M^{[p-2]}$ is expressed by the recurrent
relation \Ref{recuIR}. Hence, we conclude that for the reduced
integrable constraint, the system \Ref{systemIR} is solved by
\Ref{recuIR} without imposing any further relations on the physical
fields. The system remains completely off-shell.

\subsection{The integrable constraint}

Let us turn to the integrable constraint $K_{\scriptstyle\rm
I}$. Recall, that this is a weaker constraint than $K_{\scriptstyle\rm
IR}$ and hence gives rise to a larger spectrum. Likewise, the
dynamical equations implied by this constraint must be compatible with
the truncation \Ref{emb} since the system associated with
$K_{\scriptstyle\rm IR}$ was completely off-shell. In other words,
setting $Z_i=0=C_{ij}$ must solve all dynamical equations without
imposing further dynamics on $X_i$, $B_{ij}$.

As we have shown above, at the level $p\!=\!2$ of the supercurl
$M_{\alpha\beta}$ we find new fields arising, the explicit formulae
have been given in \Ref{MMMI}. From table \Ref{Tp2so7} we find that
$M_\bot^-$ is empty at this level, i.e.\ there are no field equations
arising at $p=2$, the projection \Ref{dyn} turns out to be satisfied
without imposing any restrictions on the level zero fields.

At level $p=4$, we expect some dynamical equations to appear in
analogy with \Ref{sc10} for the strong constraint. We discuss the
different irreducible representations of $SO(7)$, starting with the
singlet ${\bf 1}$. According to \Ref{Tp4so7}, this appears in a
particularly simple way, namely with multiplicity one. Moreover, the
table shows that
$$
\mbox{mult}\,\M_\bot^-{}\,^{[4]}_{{\bf 1}} ~=~ 1\;,
$$
i.e.\ according to \Ref{dyn}, a dynamical equation arises from the
scalar product
$$
\left\langle \M_\bot^-{}\,^{[4]}_{{\bf 1}}{}\,\right| \;
\left(TM^{[2]} - {\textstyle{2\over 3}} {\cal C}^{[4]} \right) \;,
\qquad\mbox{with}\;\;
M^{[2]}\;\;\mbox{given by \Ref{MMMI}}\;.
$$
Similarly to \Ref{sc10}, this scalar product is particularly simple
to compute because it lives on a space $\M^{[4]}_{{\bf 1}}$ of
multiplicity one. With the explicit expression from \Ref{MMMI} we arrive
at the first field equation for the enlarged system
\beq\label{dyn1}
{\cal D}^iZ_i ~=~ {\textstyle\frac43}\,\left[C_{ij},B^{ij}\right] \;.
\eeq
Note, that this equation has no analogue in the original Yang-Mills
system since in that system there is no combination of fields and
derivatives transforming as singlet of $SO(9,1)$ at this order.

Let us continue with the vector part ${\bf 7}$ which should contain
the analogue of the Yang-Mills equations of motion. For illustration,
we will describe this sector in some detail. According to the general
proceeding outlined above, we first determine the subspace $\M_\bot^-$
which by projection gives rise to the dynamical equations of the
system. It follows from \Ref{Tp4so7} and \Ref{dimn} that
$\M_\bot^-{}\,^{[4]}_{\bf 7}$ is nonempty with multiplicity one. To
determine this space explicitly, it suffices to diagonalize the
operator $S$ from \Ref{Sdef} on the space $\M_{\bot}{}_{\bf
7}^{[4]}$. A basis of the latter is e.g.~given by
\bea
(w_1)^i_{\mu\overline{\nu}\,,\,\mu_1\mu_2\rb_1\rb_2} &=&
\gamma^{mnk}_{\mu\overline{\nu}}\,\left(\gamma^{mn}_{\mu_1\mu_2}\,
\gamma^{ki}_{\rb_1\rb_2} -
\gamma^{ki}_{\mu_1\mu_2}\,\gamma^{mn}_{\rb_1\rb_2} \right) \;,
\label{bas}\\
(w_2)^i_{\mu\overline{\nu}\,,\,\mu_1\mu_2\rb_1\rb_2}
&=& \gamma^{imnk}_{\mu\overline{\nu}}\,
\left(\gamma^{mn}_{\mu_1\mu_2}\,\gamma^{k}_{\rb_1\rb_2}-
\gamma^{k}_{\mu_1\mu_2}\,\gamma^{mn}_{\rb_1\rb_2}\right)
\;,\nonumber\\
(w_3)^i_{\mu\overline{\nu}\,,\,\mu_1\mu_2\rb_1\rb_2} &=&
\gamma^{imn}_{\mu\overline{\nu}}\,\gamma^{m}_{\mu_1\mu_2}\,
\gamma^{n}_{\rb_1\rb_2}
\;,\nonumber\\
(w_4)^i_{\mu\overline{\nu}\,,\,\mu_1\mu_2\rb_1\rb_2} &=&
\gamma^{imn}_{\mu\overline{\nu}}\,\gamma^{mk}_{\mu_1\mu_2}\,
\gamma^{nk}_{\rb_1\rb_2} \;,
\nonumber
\eea
where the other components of these vectors are obtained from the
conditions \Ref{vvv0}, discussed above. Computing the action of $S$ on
this basis \Ref{bas} one finds that $\M_\bot^-{}\,^{[4]}_{{\bf 7}}$ is
spanned by
\beq\label{conv}
\M_\bot^-{}\,^{[4]}_{{\bf 7}} ~=~
\left\{w_1^i - w_2^i + 4w_3^i \right\}\;.
\eeq
The dynamical equations are finally obtained according to
\Ref{dyn} by projecting the image of $M^{[2]}_\|$ -- the latter being
entirely given by \Ref{MMMI} -- under $T$ onto the constraint vector
\Ref{conv}
$$
\left\langle \M_\bot^-{}^{[4]}_{\bf 7}{}\, \,\right| \;
\left(TM^{[2]}_\| - {\textstyle{2\over 3}} {\cal C}^{[4]} \right) \;.
$$
Explicit computation gives the following result
\bea
{\cal D}^m C_{mi} &=& {\textstyle\frac12}\,
\left[Z^m,B_{im}\right]  \;.\label{dyn7}
\eea
This gives the analogue of the Yang-Mills equations for the enlarged
system associated to the integrable dynamical constraint.  For the
strong constraint $K_{\scriptstyle\rm YM}$ this equation according to
\Ref{emb} consistently reproduces the Yang-Mills equations of motion.
Moreover, it is compatible with the absence of dynamics in the
truncation to the reduced integrable constraint $K_{\scriptstyle\rm IR}$.

Similarly, one may continue with all the other $SO(7)$
subrepresentations contained in $M_{\alpha\beta}$. As is clear from
the above proceeding, the existence of a dynamical equation first
requires the corresponding subspace $\M_\bot^-$ to be nonempty and in
addition a nontrivial projection \Ref{dyn} of the image of
$T$. Whereas validity of the first criterion may simply be extracted
from the tables collected in appendix~C, the second condition requires
a more careful calculation and has been done on the computer using
Mathematica.

We give the result for this level in linearized order, where the
complete set of dynamical equations is given by
\bea
\partial^m Z_m &=&0\;,\label{dynamic}\\[1ex]
\partial^m C_{mi} &=&0\;,\nonumber\\[1ex]
\partial_{[i}\,C_{jk]}  &=&0\;.\nonumber
\eea
The full nonlinear extensions of the first two equations have been
given in \Ref{dyn1}, \Ref{dyn7} above; likewise, the third equation
acquires nonlinear contributions, such that the field $C_{ij}$ does
not satisfy the pure Bianchi identities of a covariant field
strength. The linearized equations however are sufficient to extract
the propagating degrees of freedom contained in $Z_i$ and
$C_{ij}$. E.g.\ one of the main results of \Ref{dynamic} is the
absence of an equation in the antisymmetric ${\bf 21}$ for
$\partial_{[i}Z_{j]}$ (although the space $\M_\bot^-{}_{\,\bf 21}$ is
nonempty, the projection \Ref{dyn} has trivial image).  Whereas
$C_{ij}$ hence carries the dynamics of a propagating vector field, the
role of $Z_i$ remains somewhat unclear.

We close this section with a remark on the dynamical equations for the
higher $SO(2,1)$ spin fields at this level. Since the fields appearing
in the spectrum of $K_{\scriptstyle\rm IR}$ \Ref{spectrumIR} remain
off-shell, the only dynamical equations of higher spin can appear for
the $({\bf 3},{\bf 1})$ fields of level $p\!=\!0$ from
\Ref{spectrumI}. These fields are part of the original Yang-Mills
vector field, as such their dynamical equation is expected in the
$({\bf 3},{\bf 1})$ at level $p\!=\!4$. However, according to
\Ref{dimn} and \Ref{Tp4so7} the space $\M\,_\bot^-{}\,^{[4]}_{({\bf
3},{\bf 1})}$ is empty, such that there is no analogue of this part of the
original Yang-Mills equations of motion. The equations \Ref{dynamic}
hence contain the complete dynamical content at this level.

The task of studying the higher superfield levels which might induce
higher order equations for the fields $Z_i$ and $C_{ij}$ is left for
future work. A strong consistency check for the arising equations is
provided by their compatibility with the different truncations
\Ref{emb} to the Yang-Mills and the off-shell system, respectively.

\section{Summary}

In this paper we have analyzed the field content and the dynamical
equations induced by certain modifications of the constraint of
vanishing super curvature which is equivalent to ten dimensional
supersymmetric Yang-Mills theory. The geometric origin of the modified
constraints is a truncation of the ten-dimensional linear system,
breaking the original Lorentz symmetry $SO(9,1)$ down to
$SO(2,1)\times SO(7)$. The Lax representation thereby reduces to a
system with scalar spectral parameter \Ref{BZ} which bears strong
similarity with the Lax connection for selfdual four-dimensional
Yang-Mills theory.

Applying the general formalism of section~3, two different scenarios
have been revealed for the integrable and the reduced integrable
constraint $K_{\scriptstyle\rm I}$ and $K_{\scriptstyle\rm IR}$,
respectively. The latter induces a spectrum of $(384+384)$ fields
\Ref{spectrumIR} thereby drastically reducing the field content
\Ref{983041} of the unconstrained supercurl but still remaining
completely off-shell.

The integrable constraint $K_{\scriptstyle\rm I}$ gives rise to
additional $(31+16)$ fields which are strongly restricted by first
order differential equations which in linearized form have been given
in \Ref{dynamic}.  The complete spectrum of $SO(2,1)$ singlets thus
involves two pairs of fields ($X_i, Z_i$ and $B_{ij}, C_{ij}$) with
the same tensorial structure, but with different dimensions, since
$X_i, B_{ij}$, and $Z_i, C_{ij}$ have dimensions one and two,
respectively. The fields $X_i$ and $B_{ij}$ do not obey any equations
of motion whereas $C_{ij}$ and $Z_{i}$ appear with dynamical equations
coupled to the off-shell fields. According to \Ref{dynamic}, $C_{ij}$
contains the degrees of freedom of a propagating vector field
$\tilde{X}_i$, whereas $Z_i$ apparently is associated with an
off-shell two-form $\tilde{B}_{ij}$. Hence, we find an intriguing
duality with the original fields $X_i, B_{ij}$ which remains to be
explored in more detail.

If we reduce to four dimensions, there is a striking analogy with the
case of electromagnetism in the presence of magnetic charge (see e.g.\
ref.\cite{Zwan71}), where the field strength is build from two pieces,
a homogeneous one such that the Bianchi identity gives zero, and
another piece for which the Bianchi identity gives the magnetic
current.  Thus, we conclude that our dynamics in general involves
magnetic charges.  On the other hand, the field $B_{ij}$ has the
features of a two-form vector potential. An intriguing question is the
role of the corresponding gauge transformations $B_{ij}\to
B_{ij}+D_{[i}\Lambda_{j]}+\dots$.  However, the form of possible
interactions with higher form gauge potentials appears to be highly
restricted on general grounds (see \cite{Knae99} for a recent
discussion).  We leave all these questions to future studies.

Another point we have omitted so far, is the dynamics of the fermionic
fields of the theory which may be analyzed with exactly the same
methods that have been presented here for the bosonic sector. In
particular, the (possibly broken) supersymmetry should help to better
understand the nature of the underlying physical system.

\bigskip

In view of the geometric origin of the integrable constraints, a
natural problem is the generalization of the present approach by
studying more general reductions of the original linear system
\Ref{flat1}. Here, we have analyzed the dynamical systems associated
with the particular truncation \Ref{lam} to a one-parameter set of
light-like rays whose spatial part spans a two-dimensional plane.
Truncation to more general subvarieties may refine the
dynamics. Allowing for a more general dependence of the linear system
on the vector $v^i$ bears some striking similarities with the
harmonic superspace approach to four-dimensional $N\!=\!2$
supersymmetric Yang-Mills theory \cite{GIKOS84}.
\bigskip

We finally mention the possibility to recover in this framework and
upon dimensional reduction some of the classical higher spin gauge
theories, which have been constructed by Vasiliev (see e.g.\
\cite{Vasi96} for a review) and recently \cite{SezSun98} been brought into
the context of a possible M-theoretic origin.

\bigskip
\bigskip

\noindent
{\bf Acknowledgements}

\noindent
This work was done in part while one of us (J.-L.\ G.) was visiting
the Physics Department of the University of California at Los
Angeles. He is grateful for the warm hospitality, and generous
financial support extended to him. The work of H.\ S. was supported by
EU contract ERBFMRX-CT96-0012. Discussions with E.\ Cremmer, E.\
D'Hoker, P.\ Forgacs and K.\ Stelle have been very useful.

\bigskip
\bigskip
\bigskip

\noindent
{\Large \bf Appendix}
\begin{appendix}

\mathversion{bold}
\section{Reduction of $\sigma$-matrices.}
\mathversion{normal}

In this appendix we collect our conventions of $SO(9,1)$
$\sigma$-matrices and their decomposition into $SO(8)$ $\gamma$
matrices. We use the following particular realization:
\bea
\left(\left(\sigma^{9}\right)^{\alpha\beta}\right)&=&
\left(\left(\sigma^{9}\right)_{\alpha\beta}\right)~=~
\left(\begin{array}{cc}
-1_{8\times 8}&0_{8\times 8}\label{gamma}\\
0_{8\times 8}&1_{8\times 8}
\end{array}\right)
\;,\label{real2}\\[1ex]
\left(\left(\sigma^{0}\right)^{\alpha\beta}\right)&=&-
\left(\left(\sigma^{0}\right)_{\alpha\beta}\right)=
\left(\begin{array}{cc}
1_{8\times 8}&0_{8\times 8}\\
0_{8\times 8}&1_{8\times 8}
\end{array}\right)
\;,\nonumber\\[1ex]
\left(\left(\sigma^{i}\right)^{\alpha\beta}\right)&=&
\left(\left(\sigma^{i}\right)_{\alpha\beta}\right)=\left(\begin{array}{cc}
0&\gamma^i_{\mu \overline \nu}\\
\left(\gamma^{i\, T}\right)_{\nu\overline \mu }&0
\end{array}\right),\quad  i=1,\ldots 8
\;,\nonumber
\eea
where $\gamma^i_{\mu \overline \nu}$ denote the $SO(8)$
$\gamma$-matrices obeying
\beq
\gamma^i\gamma^{j\, T}+\gamma^j\gamma^{i\, T}~=~2\delta^{ij},\quad
i,\,j=1,\ldots, 8.
\label{o8gam}
\eeq
Our index convention here is as follows: Greek letters from the
beginning of the alphabet run from 1 to 16, letters from the middle of
the alphabet from 1 to 8, denoting the two spinor representations of
$SO(8)$. Choosing a particular eight-dimensional vector $v^i$ breaks
$SO(8)$ down to $SO(7)$ and provides a mapping between the two spinor
representations by $\gamma^{\vec v}_{\mu \overline
\nu}=v^i\gamma^i_{\mu \overline\nu}$. For notational convenience we
put $v^i=\delta^i_8$ and $\gamma^{8}_{\mu \overline \nu}=\delta_{\mu
\overline{\nu}}$, such that $\mu\leftrightarrow\bar\mu$ denotes an
$SO(7)$ covariant involution. It then follows from the Dirac algebra
\Ref{o8gam} that the matrices $\gamma^i$, $i=1,\ldots,7$ are
antisymmetric. In the main text, unless otherwise stated, Roman
indices $i,j,\dots$ from the middle of the alphabet exclusively denote
the coordinates $1,\dots,7$.

To make the $SO(2,1)$ covariance of the decomposition \Ref{gamma}
manifest, we define the action of the generators $\delta_k$,
$k=-1,0,1$ on the supercurl $M_{\alpha\beta}$ as:
\bea
\delta_k\,M_{\alpha\beta} &=& (J_k\,M+M\,J_k^T)_{\alpha\beta}+
\theta^{\alpha'}(J_k)_{\alpha'}{}^{\beta'}\,\partial_{\beta'}
\,M_{\alpha\beta}\;,
\label{so21}\\[2ex]
&&{}\mbox{with } J_0~=~\sigma^0\sigma^9\;,\quad
J_{\pm1}~=~{\textstyle\frac12}\,\sigma^8\sigma^\pm\;.\nonumber
\eea

\mathversion{bold}
\section{The $O(9,1)$ characters.}
\mathversion{normal}

In this appendix, we compute the characters of the reducible
representations which appear in section~3. The path followed is
similar to the calculation of string characters of ref.\cite{CreGer98},
and we shall refer to that paper for details. In general, the $O(9,1)$
characters are defined as $\chi(\vec v) ={\> \rm Tr \>
}\left(e^{\sum_i v_i {\cal H}_i}\right)$, where $ v_i$ are arbitrary
parameters, where the trace is taken in the representation considered,
and ${\cal H}_i$, $i=1,\ldots, 5$ are a set of five commuting elements
of the Lie algebra. Using a parametrization analogous to the one used
in ref.\cite{CreGer98} for $O(8)$ spinors, one easily sees for instance
that the character of the ${\bf 16}$ representation\footnote{In this
appendix boldface dimensions refer to $O(9,1)$ representations} is
given by
\beq
\chi_{{\bf 16}}(\vec v)=\sum_{{\epsilon_1,\cdots,\epsilon_5=\pm 1\atop
{\> \rm odd\> \#  \> }=1}}  \prod_ie^{{1\over 2}v_i\epsilon_i} \;.
\label{16ch}
\eeq

\subsection{The unconstrained character.}
In this subsection, we first compute the character associated with the
representation span by the full $M_{\alpha \beta}$, by considering the
trace over the full space ${\cal M}$. The $\alpha \beta$ indices then
contribute a factor
$$
\chi_{\mbox{\footnotesize $\underbrace{\scriptstyle {\bf 16}\otimes
{\bf 16}}_s$}}(\vec v)=
{1\over 2}\left\{
\left[\sum_{{\epsilon_1,\cdots,\epsilon_5=\pm 1\atop
{\> \rm odd\> \#  \> }=1}}  \prod_ie^{{1\over 2}v_i\epsilon_i}\right]^2
 +\sum_{{\epsilon_1,\cdots,\epsilon_5=\pm 1\atop
{\> \rm odd\> \#  \> }=1}} \prod_i e^{{1\over 2}2v_i\epsilon_i} \right\}
$$
Concerning the $\theta$ part, one works in an occupation number basis
where $N_\alpha=\theta^\alpha\partial_\alpha$ is simultaneously
diagonal for $\alpha=1,\cdots,16$. Since the Lie group generators
commute with the grading operator $\cal R $, it is convenient to
introduce in general characters of the type $\chi(\vec v|q) ={\> \rm
Tr \> }\left(q^{\cal R}e^{\sum_i v_i {\cal H}_i}\right)$. Then the
calculations becomes identical to a part of the string calculation,
where the role of ${\cal R}$ is played by the Virasoro generator
$L_0$.  Altogether, one finds that the character without any
constraint denoted $\chi_u$ is given by
\bea
\chi_u\left(\vec v|q\right)&=&
{1\over 2}\left\{
\left[\sum_{{\epsilon_1,\cdots,\epsilon_5=\pm 1\atop
{\> \rm odd\> \#  \> }=1}}  \prod_ie^{{1\over 2}v_i\epsilon_i}\right]^2
 +\sum_{{\epsilon_1,\cdots,\epsilon_5=\pm 1\atop
{\> \rm odd\> \#  \> }=1}} \prod_i e^{{1\over 2}2v_i\epsilon_i}
\right\}
\quad\times\nonumber\\[2ex]
&&{}\times\;\prod_{{\epsilon_1,\cdots,\epsilon_5=\pm 1\atop
{\> \rm odd\> \#  \> }=1}}\left(1+q\prod_i
e^{{1\over 2} v_i \epsilon_i} \right)
\label{uch}
\eea

\mathversion{bold}
\subsection{The character corresponding to ${\cal M}^\pm $.}
\mathversion{normal}

To determine them, we first compute the character with $S$
introduced. This is straightforward since $S$ is a group
invariant. Using again the occupation numbers operators $N_\alpha
=\theta^\alpha\partial_\alpha$, one may verify that
$$
\chi_S\left(\vec v|q\right)\equiv {\> \rm Tr  \> }\left(e^{\sum_{i=1}^5
v_i{\cal H}_i}S \right)
$$
\beq
=
\sum_{\mbox{\footnotesize
$\underbrace{\scriptstyle \alpha, \beta}_s$}}{\> \rm Tr  \> }_\theta
\left\{q^{\sum_\gamma
N_\gamma}\prod_i
\left(e^{v_i {\cal H}_i}\right)_{{\alpha, \beta}; {\alpha,
\beta}}  e^{v_i \sum_\rho \left({\cal H}_i\right)^\rho_{\phantom
{\delta}\rho }N_\rho
} \left(N_\alpha+N_\beta\right)\right\}
\label{chSdef}
\eeq
where the trace over $\alpha \beta$ only involves the symmetric states.
After some straightforward
computation on finds that
$$
2\chi_u\left(\vec v|q\right)-\chi_S\left(\vec v|q\right)=
\sum_{\left\{\epsilon \right\}  \left\{\epsilon' \right\}}
\prod_i e^{{1\over 2}v_i \left(\epsilon_i+ \epsilon'_i\right)}
\prod_{\left\{\eta\right\}\not=\left\{\epsilon\right\},  }\left(1+q
\prod_ie^{{1\over 2}v_i
\eta_i }\right)
$$
\beq
+\sum_{\left\{\epsilon \right\} }
\prod_i e^{v_i \left(\epsilon_i\right)}
\prod_{\left\{\eta\right\}\not=\left\{\epsilon\right\} }\left(1+q
\prod_ie^{{1\over 2}v_i
\eta_i }\right)
\label{2-S}
\eeq
Clearly,
$$
2\chi_u-\chi_S\equiv {\> \rm Tr  \> }\left(\left(S-2\right)q^{{\cal
R}}\prod_ie^{v_i{\cal
H}_i}\right) ={\> \rm Tr  \> }_{{\cal M }^+}\left(\left({\cal
R}+2\right)q^{{\cal R}} \prod_ie^{v_i{\cal H}_i}\right).
$$
An easy computation then gives
\beq
\chi^{+}\left(\vec v|q\right)\equiv {\> \rm Tr  \> }_{{\cal M }^+}
\left(q^{{\cal R}} \prod_ie^{v_i{\cal H}_i}\right)=q^{-2}
\int_0^q dx  x\left\{ 2\chi_u\left(\vec v|x\right)-
\chi_S\left(\vec v|x\right)\right\}
\label{chiZ}
\eeq
This gives
$$
\chi^{+}\left(\vec v|q)\right)=\sum_{p=0}^{14}q^p \left(p+1\right)
\sum_{\left\{\eta_1\right\}< \left\{\eta_2\right\}<\cdots <
\left\{\eta_{p+2}\right\} }e^{{1\over 2}\vec v
.\left(\sum_{r=1}^{p+2}\vec
\eta_r\right) }
$$
\beq
+\sum_{p=0}^{15}q^p \sum_{\left\{\epsilon\right\}\not=
\left\{\eta_1\right\}<
\left\{\eta_2\right\}<\cdots <
\left\{\eta_{p}\right\} }e^{{1\over 2}\vec v
.\left(\sum_{r=1}^{p}\vec
\eta_r+2\vec \epsilon\right) },
\label{chiZres1}
\eeq
where all summations over $\eta$ and $\epsilon$ are understood with an
odd total number of positive sign, respectively.  Of course the
dimensions are derived from the particular case where $\vec v=0$.  One
may verify in this way formulae given in the main text, as well as
derive others. In particular,
\beq
{\> \rm dim  \> }{\cal M^+}=16 \left(2^{17}-1\right)-
17\left(2^{16}-1\right)= 983041
\label{dim}
\eeq
It would be interesting to derive the characters $\chi^+_\|$, and
$\chi^-_\bot$ which determine the
physical field content, and the set of field equations. This is much
harder, since $K_{\scriptstyle\rm I}$ and $K_{\scriptstyle\rm IR}$
break $O(9,1)$ invariance. It is left for further studies.

\section{Group decomposition of superfields.}

In this appendix we give the decomposition of the lowest levels of the
space of superfields
\beq
\M~=~\M_\|+\M_\bot~=~\M^++\M^- \;,
\eeq
with respect to $SO(9,1)$ and $SO(2,1)\times SO(7)$, respectively.

At level $p=2$ with $\M_\|$, $\M_\bot$ defined by the strong
constraint $K_{\scriptstyle\rm YM}$ from \Ref{con10} we find the
following $SO(9,1)$ multiplicities
\beq
\mbox{
\begin{tabular}{|l||r|r|r|r|r|r|} \hline
$SO(9,1)$ mult.&{\bf 45}&{\bf 210}&{\bf 945}&{\bf 1050}&{\bf 5940}&
{\bf 6930}\\[0.1em]
\hline\hline
$\M$& 2&2&2&1&1&1\\[0.1em] \hline\hline
$\M_\|$&1&1&1&&& \\[0.1em] \hline
$\M_\bot$&1&1&1&1&1&1\\[0.1em] \hline\hline
$\M^+$&1&1&1&&1&\\[0.1em] \hline
$\M^-$&1&1&1&1&&1\\[0.1em] \hline
\end{tabular}}
\eeq
\bigskip

For the integrable constraint \Ref{KIx} this table takes the form --
now in terms of representations of $SO(2,1)\times SO(7)$
\beq\label{Tp2so7}
\begin{tabular}{|l||r|r|r|r|r|r}
\hline
$\M$ &
$4\cdot ({\bf 3},{\bf 1})$
&
$\begin{array}{c}
2\cdot ({\bf 5},{\bf 7})\\
5\cdot ({\bf 3},{\bf 7})\\
4\cdot ({\bf 1},{\bf 7})
\end{array}$
&
$\begin{array}{c}
3\cdot ({\bf 5},{\bf 21})\\
6\cdot ({\bf 3},{\bf 21})\\
6\cdot ({\bf 1},{\bf 21})
\end{array}$
&
$3\cdot ({\bf 3},{\bf 27})$
&
$\begin{array}{c}
2\cdot ({\bf 5},{\bf 35})\\
8\cdot ({\bf 3},{\bf 35})\\
4\cdot ({\bf 1},{\bf 35})
\end{array}$
&\dots\\ \hline\hline
$\M_\|$ &
$3\cdot ({\bf 3},{\bf 1})$&
$\begin{array}{c}
1\cdot ({\bf 5},{\bf 7})\\
3\cdot ({\bf 3},{\bf 7})\\
3\cdot ({\bf 1},{\bf 7})
\end{array}$
&
$\begin{array}{c}
1\cdot ({\bf 5},{\bf 21})\\
3\cdot ({\bf 3},{\bf 21})\\
4\cdot ({\bf 1},{\bf 21})
\end{array}$
&
$2\cdot ({\bf 3},{\bf 27})$
&
$\begin{array}{c}
4\cdot ({\bf 3},{\bf 35})\\
2\cdot ({\bf 1},{\bf 35})
\end{array}$
&\dots\\ \hline
$\M_\bot$ &
$1\cdot ({\bf 3},{\bf 1})$&
$\begin{array}{c}
1\cdot ({\bf 5},{\bf 7})\\
2\cdot ({\bf 3},{\bf 7})\\
1\cdot ({\bf 1},{\bf 7})
\end{array}$
&
$\begin{array}{c}
2\cdot ({\bf 5},{\bf 21})\\
3\cdot ({\bf 3},{\bf 21})\\
2\cdot ({\bf 1},{\bf 21})
\end{array}$
&
$1\cdot ({\bf 3},{\bf 27})$
&
$\begin{array}{c}
2\cdot ({\bf 5},{\bf 35})\\
4\cdot ({\bf 3},{\bf 35})\\
2\cdot ({\bf 1},{\bf 35})
\end{array}$
&\dots\\ \hline\hline
$\M^+$&
$2\cdot ({\bf 3},{\bf 1})$&
$\begin{array}{c}
1\cdot ({\bf 5},{\bf 7})\\
3\cdot ({\bf 3},{\bf 7})\\
2\cdot ({\bf 1},{\bf 7})
\end{array}$
&
$\begin{array}{c}
2\cdot ({\bf 5},{\bf 21})\\
3\cdot ({\bf 3},{\bf 21})\\
3\cdot ({\bf 1},{\bf 21})
\end{array}$
&
$2\cdot ({\bf 3},{\bf 27})$
&
$\begin{array}{c}
1\cdot ({\bf 5},{\bf 35})\\
4\cdot ({\bf 3},{\bf 35})\\
2\cdot ({\bf 1},{\bf 35})
\end{array}$
&\dots \\ \hline
$\M^-$&
$2\cdot ({\bf 3},{\bf 1})$&
$\begin{array}{c}
1\cdot ({\bf 5},{\bf 7})\\
2\cdot ({\bf 3},{\bf 7})\\
2\cdot ({\bf 1},{\bf 7})
\end{array}$
&
$\begin{array}{c}
1\cdot ({\bf 5},{\bf 21})\\
3\cdot ({\bf 3},{\bf 21})\\
3\cdot ({\bf 1},{\bf 21})
\end{array}$
&
$1\cdot ({\bf 3},{\bf 27})$
&
$\begin{array}{c}
1\cdot ({\bf 5},{\bf 35})\\
4\cdot ({\bf 3},{\bf 35})\\
2\cdot ({\bf 1},{\bf 35})
\end{array}$
&\dots \\
\hline
\end{tabular}
\eeq
The multiplicities of $\M_\|$, $\M_\bot$ here are determined by
computing the decomposition of tensor products
$$
({\bf 8}\times {\bf 8})_s \times
({\bf 8}\times {\bf 8} \times \dots \times {\bf 8})_a
\times ({\bf 8}\times {\bf 8} \times \dots \times {\bf 8})_a \;,
\qquad\mbox{etc.}\;;
$$
To obtain the multiplicities of $\M^\pm $ one needs in addition the
decomposition of the Young diagrams \Ref{M+M-}. This has been done
with the help of LiE \cite{LeCoLi92}.
\bigskip

At level $p=4$ we find with the strong constraint \Ref{con10} the
following multiplicities of the lowest dimensional $SO(9,1)$
representations
\beq\label{Tp4so10}
\mbox{\begin{tabular}{|l||r|r|r|r|r|r} \hline
&&&&& \\[-1em]
$SO(9,1)$ mult.&{\bf 10}&{\bf 120}&{\bf 126}&${\bf
\overline{126}}$&{\bf 320}
&$\;\;$\dots\\[0.1em]  \hline\hline
$\M$&   1&1&1&2&2&\dots\\[0.1em] \hline\hline
$\M_\|$& &&&1&1&\dots\\[0.1em] \hline
$\M_\bot$& 1&1&1&1&1&\dots\\[0.1em] \hline\hline
$\M^+$& &&&1&1&\dots\\[0.1em] \hline
$\M^-$& 1&1&1&1&1&\dots\\[0.1em] \hline
\end{tabular}}
\eeq
\bigskip

\noindent
whereas the integrable constraint \Ref{KIx} implies the decomposition

\beq\label{Tp4so7}
\mbox{
\begin{tabular}{|l||r|r|r|r|r|r}
\hline
$\M$&
$\begin{array}{c}
2\ti({\bf 7},{\bf 1})\\
4\ti({\bf 5},{\bf 1})\\
8\ti({\bf 3},{\bf 1})\\
1\ti({\bf 1},{\bf 1})
\end{array}$
&
$\begin{array}{c}
2\ti({\bf 7},{\bf 7})\\
11\ti({\bf 5},{\bf 7})\\
13\ti({\bf 3},{\bf 7})\\
9\ti({\bf 1},{\bf 7})
\end{array}$
&
$\begin{array}{c}
2\ti({\bf 7},{\bf 21})\\
15\ti({\bf 5},{\bf 21})\\
19\ti({\bf 3},{\bf 21})\\
14\ti({\bf 1},{\bf 21})
\end{array}$
&
$\begin{array}{c}
2\ti({\bf 7},{\bf 27})\\
7\ti({\bf 5},{\bf 27})\\
13\ti({\bf 3},{\bf 27})\\
5\ti({\bf 1},{\bf 27})
\end{array}$
&
$\begin{array}{c}
5\ti({\bf 7},{\bf 35})\\
15\ti({\bf 5},{\bf 35})\\
28\ti({\bf 3},{\bf 35})\\
10\ti({\bf 1},{\bf 35})
\end{array}$
&\dots\\[0.1em] \hline\hline
$\M_\|$&
$\begin{array}{c}
1\ti({\bf 7},{\bf 1})\\
2\ti({\bf 5},{\bf 1})\\
5\ti({\bf 3},{\bf 1})
\end{array}$
&
$\begin{array}{c}
1\ti({\bf 7},{\bf 7})\\
6\ti({\bf 5},{\bf 7})\\
7\ti({\bf 3},{\bf 7})\\
5\ti({\bf 1},{\bf 7})
\end{array}$
&
$\begin{array}{c}
7\ti({\bf 5},{\bf 21})\\
9\ti({\bf 3},{\bf 21})\\
8\ti({\bf 1},{\bf 21})
\end{array}$
&
$\begin{array}{c}
1\ti({\bf 7},{\bf 27})\\
3\ti({\bf 5},{\bf 27})\\
7\ti({\bf 3},{\bf 27})\\
2\ti({\bf 1},{\bf 27})
\end{array}$
&
$\begin{array}{c}
1\ti({\bf 7},{\bf 35})\\
5\ti({\bf 5},{\bf 35})\\
13\ti({\bf 3},{\bf 35})\\
4\ti({\bf 1},{\bf 35})
\end{array}$
&\dots\\[0.1em] \hline
$\M_\bot$&
$\begin{array}{c}
1\ti({\bf 7},{\bf 1})\\
2\ti({\bf 5},{\bf 1})\\
3\ti({\bf 3},{\bf 1})\\
1\ti({\bf 1},{\bf 1})
\end{array}$
&
$\begin{array}{c}
1\ti({\bf 7},{\bf 7})\\
5\ti({\bf 5},{\bf 7})\\
6\ti({\bf 3},{\bf 7})\\
4\ti({\bf 1},{\bf 7})
\end{array}$
&
$\begin{array}{c}
2\ti({\bf 7},{\bf 21})\\
8\ti({\bf 5},{\bf 21})\\
9\ti({\bf 3},{\bf 21})\\
6\ti({\bf 1},{\bf 21})
\end{array}$
&
$\begin{array}{c}
1\ti({\bf 7},{\bf 27})\\
4\ti({\bf 5},{\bf 27})\\
6\ti({\bf 3},{\bf 27})\\
3\ti({\bf 1},{\bf 27})
\end{array}$
&
$\begin{array}{c}
4\ti({\bf 7},{\bf 35})\\
10\ti({\bf 5},{\bf 35})\\
15\ti({\bf 3},{\bf 35})\\
6\ti({\bf 1},{\bf 35})
\end{array}$
&\dots\\[0.1em] \hline\hline
$\M^+$&
$\begin{array}{c}
1\ti({\bf 7},{\bf 1})\\
2\ti({\bf 5},{\bf 1})\\
3\ti({\bf 3},{\bf 1})
\end{array}$
&
$\begin{array}{c}
1\ti({\bf 7},{\bf 7})\\
5\ti({\bf 5},{\bf 7})\\
5\ti({\bf 3},{\bf 7})\\
3\ti({\bf 1},{\bf 7})
\end{array}$
&
$\begin{array}{c}
1\ti({\bf 7},{\bf 21})\\
6\ti({\bf 5},{\bf 21})\\
7\ti({\bf 3},{\bf 21})\\
5\ti({\bf 1},{\bf 21})
\end{array}$
&
$\begin{array}{c}
1\ti({\bf 7},{\bf 27})\\
3\ti({\bf 5},{\bf 27})\\
5\ti({\bf 3},{\bf 27})\\
2\ti({\bf 1},{\bf 27})
\end{array}$
&
$\begin{array}{c}
2\ti({\bf 7},{\bf 35})\\
6\ti({\bf 5},{\bf 35})\\
10\ti({\bf 3},{\bf 35})\\
3\ti({\bf 1},{\bf 35})
\end{array}$
&\dots\\[0.1em] \hline
$\M^-$&
$\begin{array}{c}
1\ti({\bf 7},{\bf 1})\\
2\ti({\bf 5},{\bf 1})\\
5\ti({\bf 3},{\bf 1})\\
1\ti({\bf 1},{\bf 1})
\end{array}$
&
$\begin{array}{c}
1\ti({\bf 7},{\bf 7})\\
6\ti({\bf 5},{\bf 7})\\
8\ti({\bf 3},{\bf 7})\\
6\ti({\bf 1},{\bf 7})
\end{array}$
&
$\begin{array}{c}
1\ti({\bf 7},{\bf 21})\\
9\ti({\bf 5},{\bf 21})\\
12\ti({\bf 3},{\bf 21})\\
9\ti({\bf 1},{\bf 21})
\end{array}$
&
$\begin{array}{c}
1\ti({\bf 7},{\bf 27})\\
4\ti({\bf 5},{\bf 27})\\
8\ti({\bf 3},{\bf 27})\\
3\ti({\bf 1},{\bf 27})
\end{array}$
&
$\begin{array}{c}
3\ti({\bf 7},{\bf 35})\\
9\ti({\bf 5},{\bf 35})\\
18\ti({\bf 3},{\bf 35})\\
7\ti({\bf 1},{\bf 35})
\end{array}$
&\dots\\[0.1em] \hline
\end{tabular}}
\eeq
\bigskip

\end{appendix}

\nom\modification\fin